 \documentclass[
a4paper,article, preprint, amsmath, amssymb, floatfix,
nofootinbib,wrapfloat byrevtex]{revtex4}

\usepackage{lineno,hyperref}
\modulolinenumbers[5]

 \usepackage{amsmath,mathrsfs,tabularx,graphicx,epstopdf,placeins,float}
 \usepackage{booktabs}
 \usepackage{multirow}
 \usepackage[titletoc]{appendix}
 \usepackage{titlesec}
 \usepackage{caption}
 \usepackage{color}
\usepackage{changepage}
\usepackage{graphicx}
\usepackage{subfig}
\newcolumntype{C}{>{\centering\arraybackslash}X}
\usepackage[margin=2.5cm]{geometry}

 \newcommand{\beq}[1]{\begin{equation}\label{#1}}
 \newcommand{\eeq}{\end{equation}}
 \newcommand{\bea}[1]{\begin{eqnarray}\label{#1}}
 \newcommand{\eea}{\end{eqnarray}}
 \newcommand\figcaption{\def\@captype{figure}\caption}
 \newcommand\tabcaption{\def\@captype{table}\caption}

\begin{document}
 \title{Multichannel asymmetric transmission through a dimer defect with
saturable inter-site nonlinearity}
 \author{Muhammad Abdul Wasay$^{1,2}$}
\email{wasay31@gmail.com}
\author{Magnus Johansson$^{3}$}
\email{majoh@ifm.liu.se}

\affiliation{$^1$PIP Center, Gwangju Institute of Science and Technology, Gwangju, Korea.\\
$^2$Department of Physics, University of Agriculture, Faisalabad 38040, Pakistan.\\
$^3$Department of Physics, Chemistry and Biology (IFM), Link\"{o}ping University, SE-581 83  Link\"{o}ping, Sweden}
 \begin{abstract}
We consider the asymmetric transmission properties of a Discrete Nonlinear
Schr{\"o}dinger type dimer with a saturable nonlinear intersite coupling
between
the dimer sites, in addition to a cubic onsite nonlinearity and asymmetric
linear onsite potentials. In contrast to previously studied cases with
pure onsite nonlinearities, the transmission coefficient for
stationary transmission is shown to be
a multivalued function of the {\em transmitted} intensity, in regimes of
low saturability and small or moderate transmitted intensity.  The
corresponding backward transfer map is analyzed analytically and numerically,
and shown to have either one or three distinct solutions for saturable
coupling, and zero or two solutions for the purely cubic nonlinear
coupling. As saturation strength is increased, the multi-solution regimes
disappear through bifurcations and the single-valued regime of the onsite model
is recovered. The existence of multiple solution branches yields novel
mechanisms for asymmetric left/right stationary transmission: in addition to
shifts of the positions of transmission peaks, peaks
for transmission in one direction may correspond to nonexistence of
stationary solutions propagating in the opposite direction,
at the corresponding branch and transmitted intensity. Moreover, one of these
transmission channels behaves as a nearly perfect mirror for incoming signals. The linear stability
of the stationary solutions is analyzed, and instabilities are typically
observed, and illustrated by direct numerical simulations,
in regimes with large transmission coefficient. Intersite nonlinearities are
found to
prevent the formation of a localized dimer mode in the instability-induced
dynamics.
Finally, the partial reflection and transmission of a Gaussian excitation
is studied, and the asymmetric transmission properties are
compared to previously studied onsite models.
 \end{abstract}

 \maketitle
 \smallskip

\section{Introduction}


During the last decade, there has been a major interest in using nonlinearity
for achieving non-reciprocal transmission, aiming at designing efficient wave
diodes in particular with applications within the optical domain. The basic
idea, that transmission between two waveguides with intensity-dependent (Kerr)
refractive index and different propagation constants becomes asymmetric in the
nonlinear regime, was probably first put forward by Trillo and Wabnitz
\cite{TW86}. A similar mechanism was later analyzed for a different setup
by Lepri and Casati \cite{11,lepri}, who considered the nonreciprocal
transmission through
a layered photonic (or phononic) system, with a small central segment of
nonlinear, nonmirror symmetric layers embedded in an infinite linear lattice.
In both cases, the systems were modeled using a discrete nonlinear
Schr{\"o}dinger (DNLS) equation with cubic on-site nonlinearity, with
particular focus
on the case with two nonlinear sites (DNLS dimer) which can be solved exactly.
As is well known \cite{DLS86,WS,btm}, stationary transmission through DNLS
chains with
on-site nonlinearity generically exhibits multistability and hysteresis
effects when using {\em input} intensity $|R_0|^2$ as independent parameter,
while the
existence of a backward transfer map guarantees that the stationary
transmission
coefficient is a single-valued function of the {\em output} intensity $|T|^2$ .

Later, in the context of asymmetric transmission of signals through
non-linear asymmetric dimer layers, it was shown that multistability could be
enhanced by weak saturation of the (cubic) on-site nonlinearity;
  this favors the nonreciprocal transmission and also yields opposite effects
of the rectifying action for short/long wavelength signals
\cite{assuncao,Erik}.
A similar study was carried out for saturable nonlinear oligomer DNLS segments
($N=1,2,3$)
embedded in a linear Schr{\"o}dinger chain \cite{jd}. Other relevant works
concern the asymmetric wave transmission through oligomers with cubic-quintic
on-site nonlinearity \cite{wasay18}, and the enhancement of the non-reciprocal
transmission under saturable cubic-quintic nonlinear responses for
dimers \cite{wasay3}.
 A cubic on-site nonlinearity was also shown to yield
non-reciprocal transmission if combined with an asymmetric geometric shape of
the nonlinear part \cite{LiRen}, and moreover, for a system with two nonlinear
sites separated by a number of linear sites, the transmission was shown to
be generically asymmetric unless certain resonance conditions were fulfilled
\cite{RenPRB}
(analogous resonance conditions were also obtained for the continuous
Schr{\"o}dinger equation with nonlinear $\delta$-function scatterers
\cite{RenPRB,RenPRE})

A common feature of the above mentioned earlier works is, that the nonlinearity
(cubic, cubic-quintic or saturable) resided only in the on-site terms, which
simplifies the analytical as well as the numerical treatment due to the
existence of a unique backward transfer map, which thus never yields more than
one solution for a given output $|T|$. However, although on-site nonlinearities
typically dominate in most physical applications of DNLS lattices, there are
certain situations where the additional effects of inter-site nonlinearities
may be important, e.g., for optical waveguides embedded in a nonlinear medium
\cite{Oster03}, Bose-Einstein condensates in optical lattices \cite{ST03}
(dipolar condensates in particular \cite{Serbia,Chile}), or,
more generally, when the DNLS equation is considered as a rotating-wave type
approximation of anharmonically coupled oscillators \cite{MJ06}, or as a
tight-binding approximation of the nonlinear Schr{\"o}dinger equation with
spatial periodicity in both the linear potential and nonlinearity
coefficient \cite{Abd08}. It is thus also of interest to investigate, whether
the presence of a non-negligible inter-site nonlinearity in the central
segment may have any major
qualitative effects on the non-reciprocal wave transmission. A first step in
this direction was taken in \cite{wasay2}, where transmission through an
asymmetric DNLS dimer with cubic on-site as well as inter-site nonlinearity was
studied for a special situation, assuming a particular relation between the
complex amplitudes of the two dimer sites, that allowed the derivation of
a unique backward transfer map and thus a unique solution for a given $|T|$.
However, as we will show in the present work, this is {\em not} the case for
the
generic situation in presence of inter-site nonlinearities.

In this work, we generalize the model introduced in \cite{wasay2} to have a
{\em saturable inter-site} nonlinearity between the two dimer sites,  leaving
the on-site nonlinearity cubic.
As we will see, due to the absence of a unique backward transfer map,
a small inter-site saturability typically yields {\em more
than one solution} in the regime of small $|T|$.
These solutions are distinguished by
their different relative phases between the dimer sites, and may be obtained
numerically by solving an additional equation for this phase difference.
Most importantly, the general distinctive feature of this work in view of all
previous works is that we will present the scenario of how a gradual
transition from the non-saturated to saturated case takes place, i.e., to
analyse in detail how the saturated inter-site case connects to the
non-saturated inter-site case.

Although we consider here a specific form of
saturable inter-site nonlinearities, appropriate e.g. for
photvoltaic-photorefractive materials \cite{Valley94},
the method that we present should be
applicable for transmission through generic inter-site nonlinear dimer
segments. Notably, a somewhat different form of saturable nonlinear
coupling within a dimer was recently proposed \cite{Hadad17} and
implemented through an electric circuit ladder \cite{Hadad18}.

The outline of this paper is as follows. In Sec.\ \ref{sec:Model} we introduce
the dynamical model and set up the corresponding stationary transmission
problem. We derive the corresponding backward transfer map, point out the
reason for its general non-uniqueness, and analyze its solutions
analytically in some limiting cases. In Sec.\ \ref{sec:mtc} we investigate
in more detail, by numerical means, the transitions  between regimes
with two, three, or one distinct solutions of the backward transfer map,
for increasing saturation strength. Results for the multi-channel
stationary transmission coefficients
as a function of wave number and transmitted intensity, for given parameter
values and varying saturation strength, are reported in
Sec.\ \ref{sec:transmission}. The asymmetric stationary transmission
properties are investigated in Sec.\ \ref{sec:asymmetric}, where also the
rectification factor is calculated in various regimes of saturability for
the different solution branches. Section \ref{sec:stability} reports
the linear stability analysis of the different branches of stationary
scattering solutions, with illustrations of instability-induced dynamics.
In Sec.\ \ref{sec:Gaussian} we perform dynamical simulations with
Gaussian wavepackets and discuss the transmission and rectification
properties. Concluding remarks are made in Sec.\ \ref{sec:conclusions}, and
some results for different parameter values than those used in the
main paper are shown in Appendix.

 \section{Model}
\label{sec:Model}
 We introduce the dynamical equation of the model as follows,


 \bea{}
 i\frac{dA_n}{dt}\!=\!V_nA_n\!-\!C\!\left(A_{n+1}\!+\!A_{n-1}\right)\!+\!\gamma_n|A_n|^2A_n\!+\!
\left(\!\frac{\epsilon_{n}|A_{n+1}|^2}{1\!+\!\beta|A_{n+1}|^2|A_n|^2}\!+\!\frac{\epsilon_{n-1}|A_{n-1}|^2}{1\!+\!\beta|A_{n-1}|^2|A_n|^2}\!\right)\!A_n .
 \label{dynamical}
 \eea
In Eq.\ \eqref{dynamical}, $n$ is the lattice site counter,
$A_n$ is the amplitude at site $n$, and $V_n$ is the linear on-site energy of
each
site inside the one-dimensional lattice. The parameter $\gamma_n$ determines
the strength of the local nonlinearity,
which we take to have a standard cubic (Kerr) form, while
$\epsilon_n$ represents the nonlocal (inter-site) saturable nonlinearity.
We saturate only the inter-site nonlinearity as the effect of saturating the
on-site cubic nonlinearity has been addressed in earlier work
\cite{jd,assuncao,wasay3,Erik}. $\beta$ is the saturation parameter;
in the unsaturated limit $\beta =0$ we recover the DNLS model with
cubic inter-site
nonlinearity studied in \cite{wasay2} (relevant e.g.\ for dipolar Bose-Einstein
condensates \cite{Chile}), while in the strongly saturated limit
$\beta \rightarrow \infty$ the coupling becomes governed only by the linear
coupling constant $C$, and the model reduces to the standard cubic on-site
DNLS model \cite{11,lepri}.
Like the standard DNLS model \cite{EJ03}, the system is Hamiltonian,
with the saturable inter-site terms arising from additional terms
$\frac{\epsilon_n}{\beta} \ln \left(1+\beta|A_n|^2|A_{n+1}|^2\right)$
in the Hamiltonian.
Without loss of generality, the parameter $C$
will be chosen to be unity. Focusing our attention on a dimer
situated at lattice sites 1 and 2 with nonlinear inter-site interactions
only between the two dimer sites, the site
dependent parameters $\gamma$ and $V$ have non-zero
contributions only in the
region $1\leq n\leq2$, and $\epsilon$ only for $n=1$.
This means that waves can propagate freely outside the dimer.

The set of dynamical equations \eqref{dynamical} has stationary solutions of
the form $A_n(t)=A_ne^{-i\omega t}$.
When a signal (incoming or outgoing wave) is away from the dimer, the system
is linear, and these solutions satisfy the dispersion relation
$\omega=-2$cos$k$, with $k$ being the wave vector of some specific harmonic
component of the wave and $0\leq k\leq\pi$. These solutions render
Eq.\ \eqref{dynamical} stationary. The resulting stationary equation can be
written
in the form of a backward-map analogous to \cite{11,btm},
 \bea{}
A_{n-1}=-A_{n+1}+\left(V_n-\omega+\gamma_n|A_n|^2+\frac{\epsilon_{n}|A_{n+1}|^2}{1+\beta|A_{n+1}|^2|A_n|^2}+\frac{\epsilon_{n-1}|A_{n-1}|^2}{1+\beta|A_{n-1}|^2|A_n|^2}\right)A_n .
\label{stationary}
 \eea
In the absence of inter-site nonlinearities ($\epsilon_n \equiv 0$), this
relation allows one to immediately
construct the amplitudes by a backward iteration, assuming that the solution
is known at $n\rightarrow \infty$. By contrast, with nonzero $\epsilon_1$
an additional relation between the complex amplitudes
$A_1$ and $A_2$ at the nonlinear dimer sites is needed.
 We will focus on the scattering properties of plane wave solutions of the
following form,
\bea{}
A_n=
\Bigg\{
  \begin{array}{c}
  R_0 e^{ikn}+Re^{-ikn}             \qquad n\leq 1
  \\
  Te^{ikn}                          ~~\qquad\qquad\qquad n\geq 2 \\
  \end{array} ,
  \label{planewave}
\eea
 where $R_0$, $R$ and $T$ are the amplitudes of incoming, reflected and
transmitted wave, respectively. Applying the ansatz in Eq.\ \eqref{planewave}
site-by-site,  we get at site $n=0$,
 \bea{}
 A_0=R_0+R ,
 \label{site0}
 \eea
  and at site $n=1$,
 \bea{}
 A_1=R_0e^{ik}+Re^{-ik} .
 \label{site1}
 \eea
 With $A_0$ and $A_1$, the amplitudes of the reflected and incident waves can
be computed as
\bea{}
R=\frac{A_0 e^{ik}-A_1}{e^{ik}-e^{-ik}} ,
\label{R}
\eea
and
\bea{}
R_0=\frac{A_0 e^{-ik}-A_1}{e^{-ik}-e^{ik}} .
\label{R0}
\eea

 We can rewrite the backward map \eqref{stationary} for the dimer
($n=2$), with the inter-site nonlinear interactions considered only between
the two dimer sites, as
\bea{}
A_1=-A_3+\left(V_2-\omega+\gamma_2|A_2|^2+\frac{\epsilon_1|A_{1}|^2}{1+\beta|A_{1}|^2|A_2|^2}\right)A_2 .
\label{n2}
\eea
With Eq.\ \eqref{planewave}, the wave amplitudes at the dimer interface
for the outgoing, right-propagating waves ($k>0$) is
$A_2=Te^{2ik}$ and $A_3=Te^{3ik}$. In addition, we may obtain an expression for
$|A_1|^2$ from the general current conservation law for stationary solutions,
\bea{}
\textmd{Im}[A_n^\ast A_{n+1}]=|T|^2 \sin k,
\label{cl}
\eea
for all $n$.
Without loss of generality, we may choose the arbitrary overall phase
such that the amplitude $A_1$ at site 1 is real.
From \eqref{cl} with $n=1$ and \eqref{planewave} with $n=2$ it then follows
straightforwardly that $A_1 \textmd{Im}(Te^{2ik}) = |T|^2 \sin k$, i.e.,
\bea{}
|A_1|^2=\frac{|T|^4\textmd{sin}^2k}{[\textmd{Im}(Te^{2ik})]^2} .
\label{f1}
\eea
The fact that $T$ is not real in general then leads to the following relation,
\bea{}
|A_1|^2=\frac{|T|^2\sin^2k}{[\sin(2k+\varphi)]^2},
\eea
where $\varphi=\textmd{arg}(T)$. Thus, Eq.\ \eqref{n2} becomes
\bea{}
A_1=-Te^{3ik}+\left(V_2-\omega+\gamma_2|T|^2+\frac{\epsilon_1|T|^2\textmd{sin}^2k}{\left[\textmd{sin}(2k+\textmd{arg}(T))\right]^2+\beta|T|^4\textmd{sin}^2k}\right)Te^{2ik} ,
\label{n3}
\eea
which can be re-written as
 \bea{}
 A_1=Te^{2ik}(\delta_2-e^{ik}) ,
 \label{n5}
 \eea
 with $\delta_2=V_2-\omega+\gamma_2|T|^2+\frac{\epsilon_1|T|^2\textmd{sin}^2k}{\left[\textmd{sin}(2k+\textmd{arg}(T))\right]^2+\beta|T|^4\textmd{sin}^2k}$.

 Now in a similar way, for $n=1$ with the nonlinear inter-site interactions
only between the two dimer sites, from Eq.\eqref{stationary} we get
\bea{}
A_0=-A_2+\left(V_1-\omega+\gamma_1|A_1|^2+\frac{\epsilon_1|A_{2}|^2}{1+\beta|A_1|^2|A_{2}|^2}\right)A_1 ,
\label{n21}
\eea
which together with Eq.\eqref{n5} leads to
%
\bea{}
A_0=-Te^{2ik}+\left(V_1-\omega+\gamma_1|T|^2|\delta_2-e^{ik}|^2+\frac{\epsilon_1|T|^2}{1+\beta|T|^4|\delta_2-e^{ik}|^2}\right)
Te^{2ik}(\delta_2-e^{ik}) ,
\label{n31}
\eea
which can be rewritten as
\bea{}
A_0=Te^{2ik}\left[\delta_1(\delta_2-e^{ik})-1\right] ,
\label{n6}
\eea
%
with $\delta_1=V_1-\omega+\gamma_1|T|^2|\delta_2-e^{ik}|^2+\frac{\epsilon_1|T|^2}{1+\beta|T|^4|\delta_2-e^{ik}|^2}$.

 This implies that the transmission coefficient $t(k,|T|^2; \arg(T))$ can now
be computed by using Eq.\ \eqref{n6} and Eq.\ \eqref{n5} in Eq.\ \eqref{R0}.
The result is
\bea{}
t(k,|T|^2;\arg(T))=\frac{|T|^2}{|R_0|^2}=\left|\frac{e^{-ik}-e^{ik}}{(\delta_2-e^{ik})(\delta_1-e^{ik})-1}\right|^2 .
\label{t}
\eea
 For the left-propagating waves (with $k<0$), a similar computation yields
the transmission coefficients, i.e., we only need to exchange the
subscripts 1 \& 2.

Note that, in contrast to previously studied DNLS-type transmission problems
(e.g., \cite{11,lepri,DLS86,WS,btm,assuncao,Erik,jd,wasay18,wasay3,LiRen,RenPRB}),
Eq.\ \eqref{t} does {\em not} necessarily determine the transmission
coefficient for a plane wave of wave vector $k$ as a unique function of
the transmitted intensity $|T|^2$, since in presence of intersite
nonlinearities, there may be multiple solution
branches $i$ corresponding to the same $|T|^2$ but with different phases
$\varphi_i\equiv \arg(T)$. Since Eq.\ \eqref{n3} was obtained by assuming
$A_1$ real,
the interpretation of the quantity ``$\arg(T)$" is an additional phase shift
between the two dimer sites 1 and 2, due to the internal properties of the
dimer. The computation of this additional phase factor $\arg(T)$ is
nontrivial. Imposing the "reality'' assumption on $A_1$ by putting the
imaginary part of the right-hand side of Eq.\ \eqref{n3} to zero,
leads to the following equation,
 \bea{}
 \frac{\sin\left[3k+\arg(T)\right]}{\sin\left[2k+\arg(T)\right]}=V_2-\omega+\gamma_2|T|^2+\frac{\epsilon_1|T|^2\sin^2k}{\left[\sin(2k+\arg(T))\right]^2+\beta|T|^4\sin^2k}.
 \label{argT}
 \eea
Solving Eq.\eqref{argT} analytically for $\arg(T)=\varphi_i(|T|,k)$ is
nontrivial in the general case, but we may immediately note that adding
any multiple of $\pi$ to a solution gives another solution, so we may restrict
to the interval $0\leq \varphi < \pi$ (adding $\pi$ just switches an overall
sign).
For the case with pure on-site nonlinearity ($\epsilon_1=0$), we recover the
unique solution
 \bea{}
 \varphi_{\epsilon_1=0}(|T|,k)  = -2k + \arctan\left[\frac{\sin k}
{V_2+\cos k+\gamma_2|T|^2}\right].
 \label{argT1}
 \eea
Note that in the linear limit ($|T|\to 0$) and non-scattering case
($V_2 = 0$), Eq.\ \eqref{argT1} yields
$\arg(T)=-k$ (due to the choice of origin in  \eqref{planewave}).

With unsaturated inter-site nonlinearity ($\epsilon_1\neq 0$ but $\beta=0$),
Eq.\ \eqref{argT} can be written as a quadratic equation for
$y\equiv \sin^2(2k+\varphi)$:
 \bea{}
 y^2(e^2+\sin^2k)+y (2e \epsilon_1|T|^2-1) \sin^2k  + \epsilon_1^2|T|^4 \sin^4k = 0,
 \label{quadratic}
 \eea
where $e\equiv V_2+\cos k + \gamma_2 |T|^2$.
Thus, it is clear from \eqref{quadratic} that for unsaturated inter-site
nonlinearities
there are generally {\em two distinct
solutions} for {\em small} $|T|$, and {\em no} solutions for {\em large} $|T|$.
For small $|T|$ we may express the solutions to $O(|T|^4)$ as
\bea{}
\varphi_1 \simeq -2k +  \arctan\left[\frac{\sin k}{V_2+\cos k}
\left\{1-\frac{\gamma_2+\epsilon_1[(V_2+\cos k)^2+\sin^2k]} {V_2+\cos k}|T|^2
\right\}
\right],
 \label{unsat1}
 \eea
and
\bea{}
\varphi_2 \simeq -2k ,
 \label{unsat2}
 \eea
respectively. Thus we note that the solution $\varphi_1$ in \eqref{unsat1}
coincides with the small-$|T|$ limit of the pure on-site solution \eqref{argT1}
when $\epsilon_1 \rightarrow 0$, while the solution $\varphi_2$ in
\eqref{unsat2} yields a new possible transmission channel not existing for
pure on-site nonlinearity (the existence of this additional channel was suppressed in \cite{wasay2} due to the choice of a specific relation between the complex amplitudes of the dimer sites).

For the general case with saturable inter-site nonlinearity
($\beta \neq 0$), Eq.\ \eqref{argT} instead becomes a cubic equation for $y$
($0\leq y \leq 1$),
 \bea{}
(y+\beta |T|^4\sin^2 k)^2 (1-y) \sin^2 k = y
\left[e(y+\beta|T|^4 \sin^2 k)+\epsilon_1|T|^2 \sin^2 k \right]^2 .
 \label{cubic}
 \eea
Thus, depending on the parameter values, there are {\em either one or three}
distinct solutions $\varphi_i$, corresponding to different possible
transmission channels.
 By analyzing the coefficients in \eqref{cubic}, we find that in the limit of
small $|T|$ there is a transition at $\beta/\epsilon_1^2 =1/4$, so that for
$\beta/\epsilon_1^2 > 1/4$ there is only one real and positive solution for
$y$,
corresponding to
the phase factor $\varphi_1$ in \eqref{unsat1} for the unsaturated case.
On the other hand, for $\beta/\epsilon_1^2 < 1/4$ two additional solutions
appear, both originating from the solution $\varphi_2$ in \eqref{unsat2}
for the unsaturated
case, with corrections of order $|T|^4$ and higher. Below,
the value $\beta/\epsilon_1^2 = 1/4$ will be taken as indicating the
transition between regimes of ``low saturation'' and ``medium saturation''.

Moreover, in the limit of large  $|T|$, $y=0$ is the only real solution,
thus yielding a single channel with phase factor $\varphi_2$ given by \eqref{unsat2}. Note that this agrees also with the large-$|T|$ limit of the pure on-site
solution \eqref{argT1}, as it should since the coupling becomes effectively
linear due to strong saturation.
For general nonzero $|T|$, the phase factor $\varphi_i(|T|,k)$ is computed
numerically,
and we will illustrate the typical scenario for different regimes
of saturability in the following section. As we will see, there are
significant regimes with three distinct channels for small and intermediate
values
of $|T|$ when $\beta$ is not too large.

\section{Multiple transmission channels}
\label{sec:mtc}


As described above, one of our goals is to investigate how the saturated case
connects to the non-saturated case, which hinges on the detailed investigation
of scenarios occurring at different levels of saturation. The analysis is thus
divided into
three distinct saturation regimes with carefully chosen representative
saturation values. To be specific, we fix the value of $\epsilon_1$ to
$\epsilon_1=0.5$, consider
an ``ultra-low saturation'' regime for
$\beta=0.01$ (i.e., $\beta/\epsilon_1^2 =0.04 \ll 1/4$), a
low saturation regime for $\beta=0.05$
(i.e., $\beta/\epsilon_1^2 =0.2 < 1/4$),  and a medium saturation
regime $\beta=0.5$ (i.e., $\beta/\epsilon_1^2 =2 > 1/4$) (the regime
of stronger saturations is less interesting since it essentially reproduces
well known results for pure on-site nonlinearity \cite{11,lepri}).
Unless otherwise noted, we will also fix the on-site nonlinearity strength
on the dimer symmetric as $\gamma_1=\gamma_2=1$, and express the asymmetric
on-site potential as $V_{1,2}=V^{(0)}(1\pm\varepsilon_V)$, with specifically
chosen $V^{(0)}=-2.50$ and $\varepsilon_V=0.05$
(some results for different values
of $\varepsilon_V, V^{(0)}$ are discussed in Appendix \ref{AppB}).

  \begin{figure}[!htbp]
  \begin{minipage}[h]{0.23\linewidth}
    \centering
    \includegraphics[width=\linewidth]{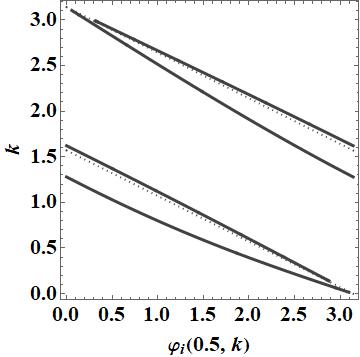}
  \end{minipage}
  \begin{minipage}[h]{0.23\linewidth}
    \centering
    \includegraphics[width=\linewidth]{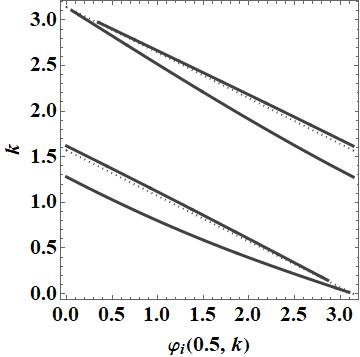}
  \end{minipage}
  \begin{minipage}[h]{0.23\linewidth}
    \centering
    \includegraphics[width=\linewidth]{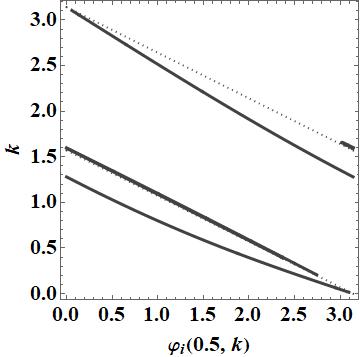}
  \end{minipage}
  \begin{minipage}[h]{0.23\linewidth}
    \centering
    \includegraphics[width=\linewidth]{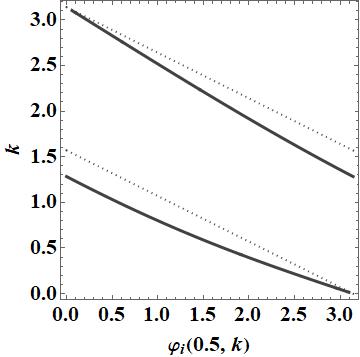}
  \end{minipage}
  \\
  \begin{minipage}[h]{0.23\linewidth}
    \centering
    \includegraphics[width=\linewidth]{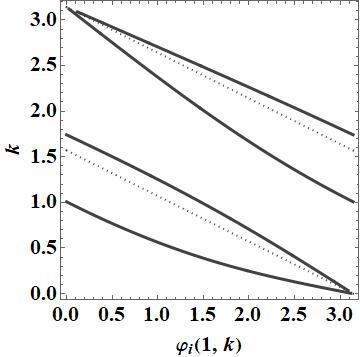}
  \end{minipage}
  \begin{minipage}[h]{0.23\linewidth}
    \centering
    \includegraphics[width=\linewidth]{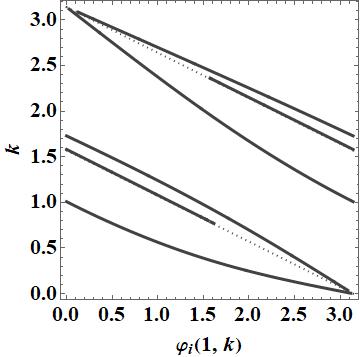}
  \end{minipage}
  \begin{minipage}[h]{0.23\linewidth}
    \centering
    \includegraphics[width=\linewidth]{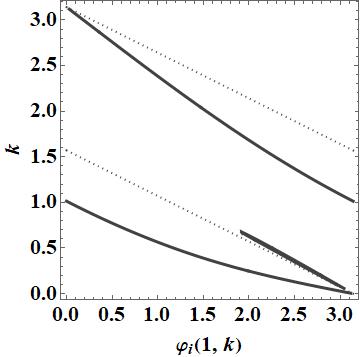}
  \end{minipage}
  \begin{minipage}[h]{0.23\linewidth}
    \centering
    \includegraphics[width=\linewidth]{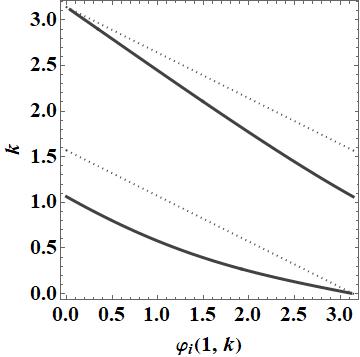}
  \end{minipage}
  \caption{Contour plot of solutions of Eq.\eqref{argT} as a function of $k$
and $\arg(T)=\varphi_i(|T|,k)$ for $V_2=-2.3750$, $\gamma_2=1$ and
$\epsilon_1=0.5$. Upper row:  $|T|=0.5$;
lower row: $|T|=1$. Columns from
left to right:  $\beta=0$,  $\beta=0.01$,  $\beta=0.05$,  $\beta=0.5$. }
 \label{fargTb0}
  \end{figure}
\begin{figure}[!htbp]
  \begin{minipage}[h]{0.24\linewidth}
    \centering
    \includegraphics[width=\linewidth]{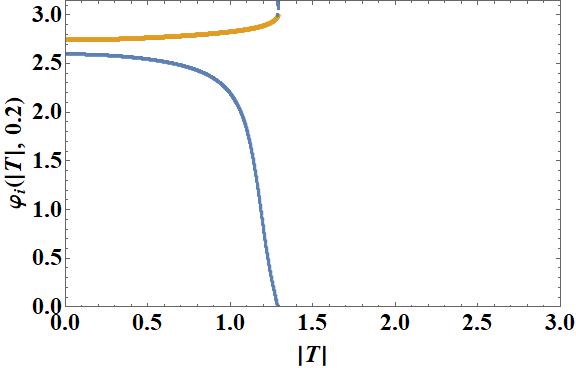}
  \end{minipage}
    \begin{minipage}[h]{0.24\linewidth}
    \centering
    \includegraphics[width=\linewidth]{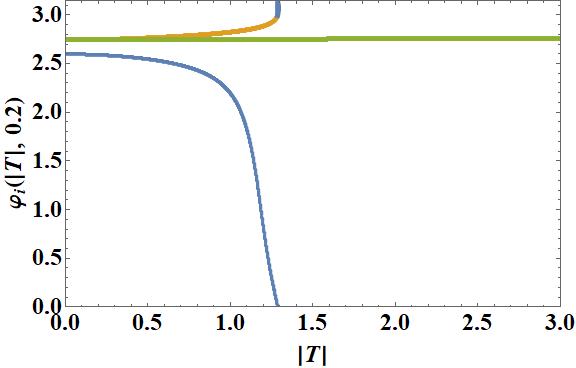}
  \end{minipage}
    \begin{minipage}[h]{0.24\linewidth}
    \centering
    \includegraphics[width=\linewidth]{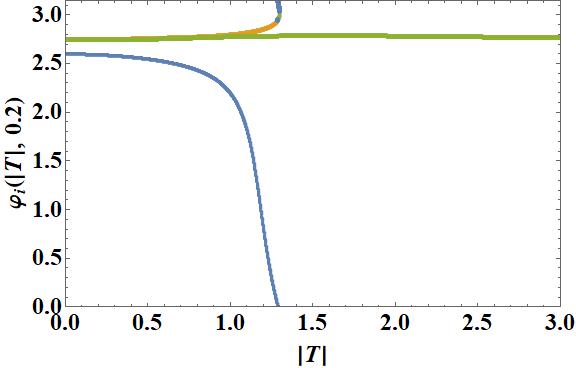}
  \end{minipage}
    \begin{minipage}[h]{0.24\linewidth}
    \centering
    \includegraphics[width=\linewidth]{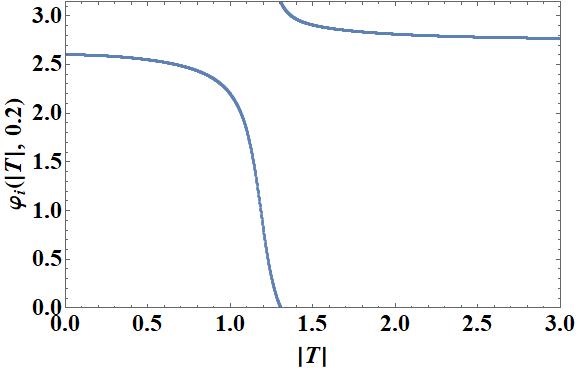}
    \end{minipage}
    \\
   \begin{minipage}[h]{0.24\linewidth}
    \centering
    \includegraphics[width=\linewidth]{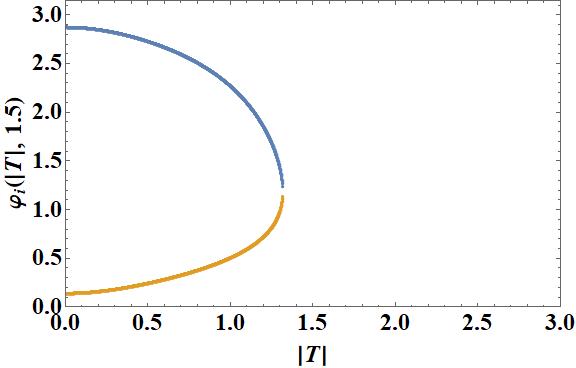}
  \end{minipage}
   \begin{minipage}[h]{0.24\linewidth}
    \centering
    \includegraphics[width=\linewidth]{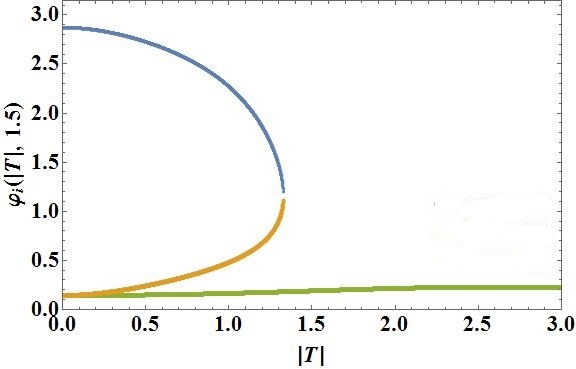}
  \end{minipage}
  \begin{minipage}[h]{0.24\linewidth}
    \centering
    \includegraphics[width=\linewidth]{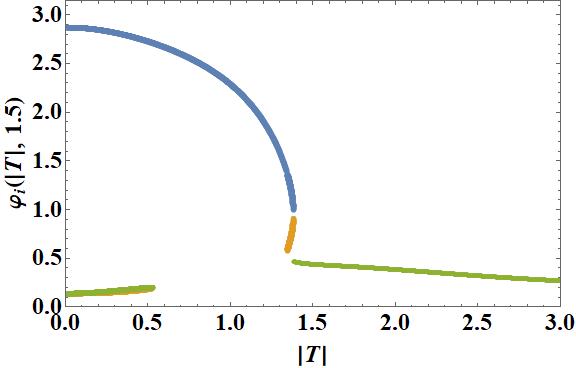}
  \end{minipage}
  \begin{minipage}[h]{0.24\linewidth}
    \centering
    \includegraphics[width=\linewidth]{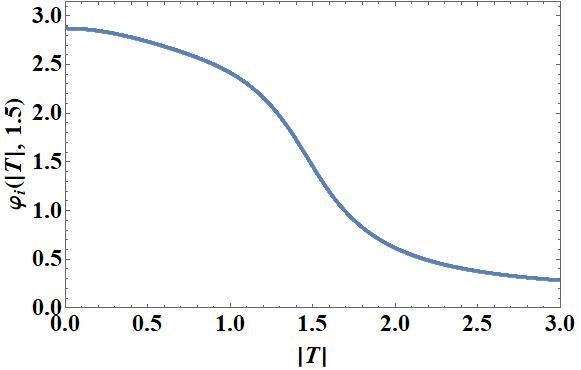}
  \end{minipage}
  \\
  \begin{minipage}[h]{0.24\linewidth}
    \centering
    \includegraphics[width=\linewidth]{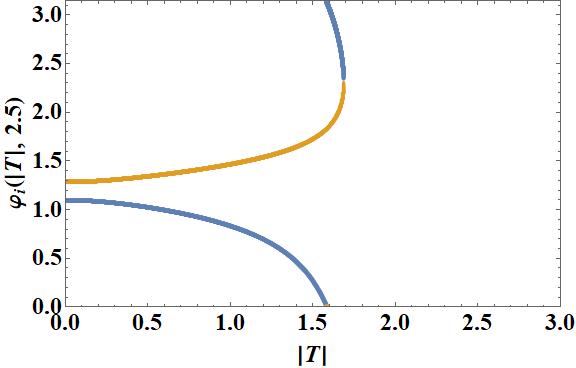}
  \end{minipage}
  \begin{minipage}[h]{0.24\linewidth}
    \centering
    \includegraphics[width=\linewidth]{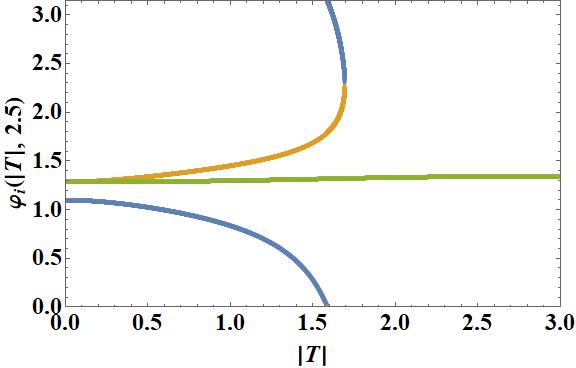}
  \end{minipage}
  \begin{minipage}[h]{0.24\linewidth}
    \centering
    \includegraphics[width=\linewidth]{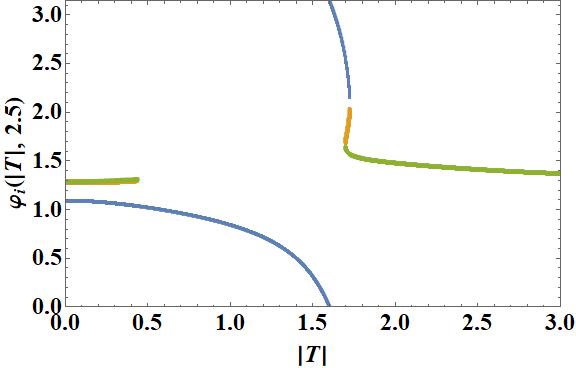}
  \end{minipage}
  \begin{minipage}[h]{0.24\linewidth}
    \centering
    \includegraphics[width=\linewidth]{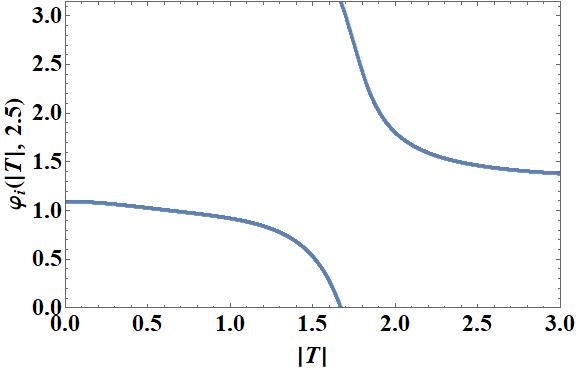}
  \end{minipage}
  \caption{Plots of $\varphi_i(|T|,k)$ as a function of $|T|$ with three fixed
    values of $k$: $k=0.2$ (upper row), $k=1.5$
    (middle row), $k=2.5$
    (bottom row). Columns from left to right:
      unsaturated case ($\beta=0$),
ultra-low ($\beta=0.01$), low ($\beta=0.05$), and medium saturation
($\beta=0.5$) strengths. All other parameter values as before. Blue, orange
and green curves correspond to the first, second and third solution branch,
respectively.}
\label{multisols}
  \end{figure}

The solutions to Eq.\ \eqref{argT} for two distinct nonzero
$|T|$ values ($|T|$=0.5 and $|T|$=1) in the unsaturated and the three
saturation regimes are illustrated in Fig.\ \ref{fargTb0} for $0\leq k\leq\pi$
versus $0\leq\varphi_i(|T|,k)\leq\pi$. The ''dotted'' lines in these
figures represent the singularity in the left-hand side of  \eqref{argT}
at $y=0$ ($\varphi = -2k \mod \pi$), which lies very close to the actual
solution for some cases.
In Fig.\ \ref{multisols}, the corresponding solutions ($\varphi_i(|T|,k)$) are
presented  as a function of $|T|$ for various (fixed) values of $k$.

Firstly, for the non-saturated case ($\beta=0$, left columns in
Figs.\ \ref{fargTb0}-\ref{multisols}), as predicted from \eqref{quadratic}
there is a two-solution regime which exists for all $k$ for small $|T|$,
and no valid solutions for higher $|T|$ above a certain cut-off, which differs
slightly for different $k$.
For the chosen set of parameter values, in the case of relatively small
$k\leq1.4$ this cut-off value is approximately $|T|=1.2$,
and slightly increases for larger $k$ values, i.e., for $2.8\leq k\leq\pi$, the cut-off is at $|T|=1.8$ approximately.
Thus, there are either two or zero transmission channels in the case of zero
saturation.
%

   As soon as we switch from the unsaturated to the ultra-low saturation
(second column in  Figs.\ \ref{fargTb0}-\ref{multisols}),
a third solution immediately appears in the small-$|T|$ solution regime,
as predicted from \eqref{cubic}. Note that this additional solution
(green curve in Fig.\ \ref{multisols}) as expected almost coincides with the
``singularity line'' $\varphi = -2k \mod \pi$ in Fig.\ \ref{fargTb0}.
Above a certain cut-off depending upon $k$
(which is essentially the same as for the unsaturated case),
only this new third solution persists throughout the parameter space.
So there is either a three-solution regime (for small $|T|$) or a
single-solution regime for higher $|T|$.


For the case of low saturations ($\beta=0.05$, third column in
Figs.\ \ref{fargTb0}-\ref{multisols}), the three-solution regime as predicted
always persists for small $|T|$ (orange and green branches in
Fig.\ \ref{multisols} almost coincide) and also for small $k$ (the
apparent gap close to $k=0$ in Fig.\ \ref{fargTb0} is due to graphics
limitations).
However, for slightly larger $k$ ($k>0.6$), the scenario is different as
compared to the ultra-low case:  the three-solution regime at small $|T|$
is interrupted by a single-solution regime
(blue curve only in Fig.\  \ref{multisols}) for some interval of $|T|$, and
then there is a  second (small) three-solution regime which exists upto
a cut-off as in the ultra-low case. For all higher $|T|$, there is a
single solution (the third solution, green curve in Fig.\  \ref{multisols}).
Note however that the single-solution regime sandwiched between the two
three-solution regimes \textit{does not} belong to the
'third solution branch', but rather it belongs to the 'first solution branch'.

Finally, in the medium saturation range ($\beta=0.5$, right
  column in
Figs.\ \ref{fargTb0}-\ref{multisols})), the multi-solutions are
entirely suppressed leaving behind only a single-solution regime for all $k$
and $|T|$.





Also, it is to be noted (see Appendix \ref{AppB})
that the stretch of intensities that exhibit the
multi-channel regime strongly depends on the energy at site 2, represented by
the parameter $V_2$. For a smaller $V_2$ the multi-channel regime persists for
a longer stretch of intensities (i.e., the cut-off is higher) and vice versa.

\section{Effects of saturation on multi-channel transmission}
\label{sec:transmission}

\begin{figure}[!htbp]
  \begin{minipage}[h]{0.32\linewidth}
    \centering
    \includegraphics[width=\linewidth]{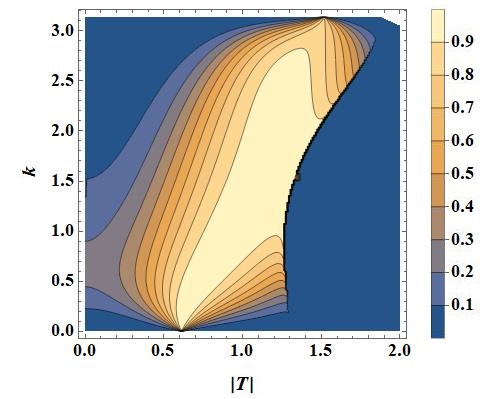}
  \end{minipage}
  \hspace{-0.2cm}
  \begin{minipage}[h]{0.32\linewidth}
    \centering
    \includegraphics[width=\linewidth]{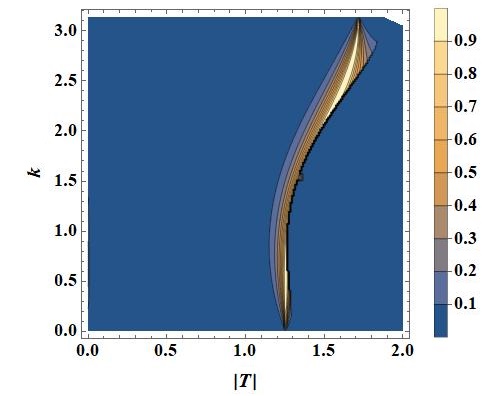}
  \end{minipage}
  \hspace{-0.2cm}
  \begin{minipage}[h]{0.34\linewidth}
    \centering
    \includegraphics[width=\linewidth]{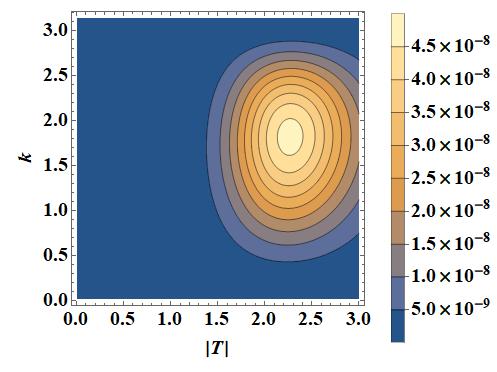}
  \end{minipage}
  \begin{minipage}[h]{0.32\linewidth}
    \centering
    \includegraphics[width=\linewidth]{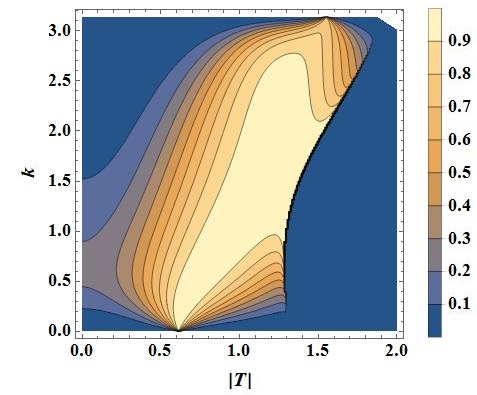}
  \end{minipage}
  \begin{minipage}[h]{0.32\linewidth}
    \centering
    \includegraphics[width=\linewidth]{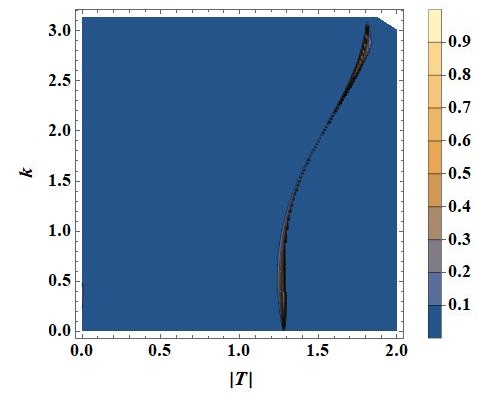}
  \end{minipage}
  \begin{minipage}[h]{0.32\linewidth}
    \centering
    \includegraphics[width=\linewidth]{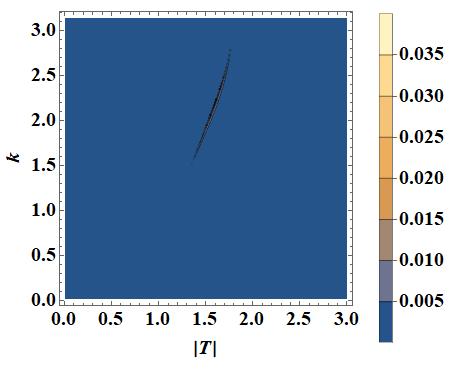}
  \end{minipage}
  \begin{minipage}[h]{0.32\linewidth}
    \centering
  \includegraphics[width=\linewidth]{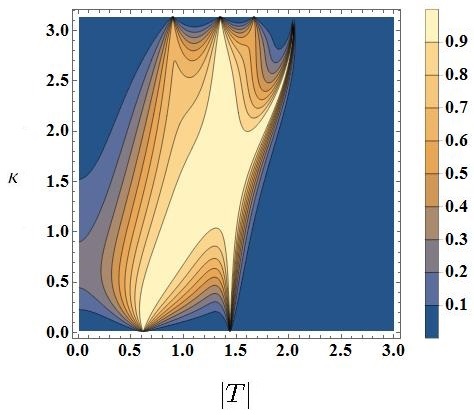}
  \end{minipage}
  \caption{Color plots of transmission coefficient as a function of
$k$ and $|T|$ for the three different saturation regimes.
Upper row: $\beta=0.01$; middle row: $\beta=0.05$; lower row:
$\beta=0.5$. Left, middle and right panels in two upper rows correspond to
the 1st, 2nd and 3rd transmission channel, respectively. In all figures,
$\epsilon_{1}=0.5$, $\varepsilon_V=0.05$, $V^{(0)}=-2.5$, and $\gamma_{1,2}=1$.
}
 \label{u00001}
   \end{figure}

In this section, we present in Fig.\  \ref{u00001} the transmission scenario
via density plots of the
transmission coefficient $t(k,|T|^2)$ from Eq.\ \eqref{t}, for the parameter values
corresponding to the different
saturation regimes discussed above. The first, second and third transmission
channels correspond to blue, orange and green curves from
Fig.\ \ref{multisols},
respectively. The results for the two transmission channels in the
unsaturated case ($\beta=0$) are visually identical to those of the
first and second channels for the ultra-low saturation ($\beta=0.01$),
and thus we do not show these figures below.
%

\subsubsection*{Ultra Low Saturation Regime, $\beta=0.01$}



From the upper row in Fig.\ \ref{u00001}, we see that the major regimes
of good transmission occur along the first channel, and that transmission
along the third channel is essentially negligible. Note however that there
is a narrow peak with close to perfect transmission also along the second
channel, most apparent around $k=0.2$ and $k=2.5$. This second peak appears
also for pure on-site nonlinearities \cite{11,wasay18}; what is important
to note here is thus that with inter-site nonlinearity (unsaturated or with
ultra-low saturation), the second transmission peak moves over to the second
branch of solution, and thus there is only one peak in each of the two first
channels.


%

\subsubsection*{Low Saturation Regime, $\beta=0.05$}

Here, we see from the second row of Fig.\ \ref{u00001} that the transmission
along the first channel is essentially the same as for smaller saturation,
but that of the second channel narrows down as the existence region for the
corresponding solution shrinks, as discussed above.
Note that when $\beta=0.05$ there are two distinct three-solution regimes
for $k>0.6$ (see Fig.\ \ref{multisols}), but the
transmssion along the second and third channels is always negligible in the
small-$|T|$ region. Note also that since the third solution now connects
to the second solution when $k>0.6$, it also picks up some noticeable, but
still small, transmission close to the connection point (seen for
$1.4\lesssim |T| \lesssim 1.8$ in Fig.\ \ref{u00001}).


%


%

\subsubsection*{Medium Saturation Regime, $\beta=0.5$}

As discussed above, in the medium saturation regime
only single solutions persist
throughout the parameter space.
The transmission coefficient shown in the lowest part of Fig.\  \ref{u00001}
is qualitatively similar to analogous plots for the pure on-site nonlinearity
case \cite{11,wasay18}: two separate transmission peaks for small $k$ which
merge to a single, broad peak around $k=\pi/2$, and then split up again,
eventually yielding four distinct peaks close to $k=\pi$. Note also that
although a propagating solution exists for arbitrarily large $|T|$ and all $k$,
the transmission coefficient is essentially negligible for $|T|\gtrsim 2.1$.

%


 \section{Asymmetric multi-channel transmission}
\label{sec:asymmetric}

\begin{figure}[!htbp]
  \begin{minipage}[h]{0.32\linewidth}
    \centering
    \includegraphics[width=\linewidth]{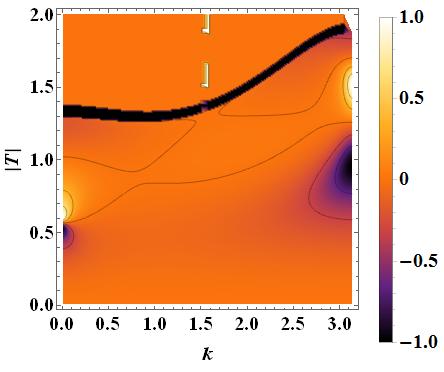}
  \end{minipage}
  \begin{minipage}[h]{0.32\linewidth}
    \centering
    \includegraphics[width=\linewidth]{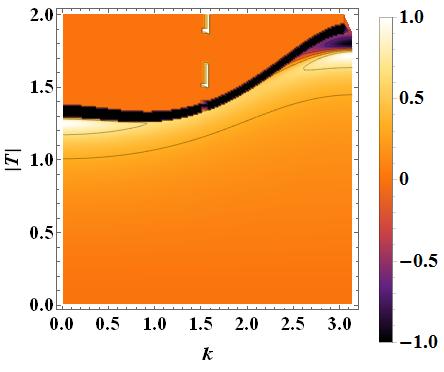}
  \end{minipage}
  \begin{minipage}[h]{0.32\linewidth}
    \centering
    \includegraphics[width=\linewidth]{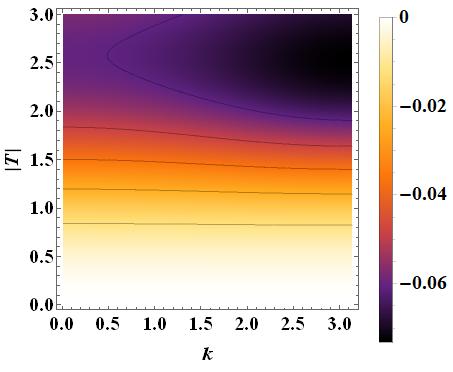}
  \end{minipage}
  \begin{minipage}[h]{0.32\linewidth}
    \centering
    \includegraphics[width=\linewidth]{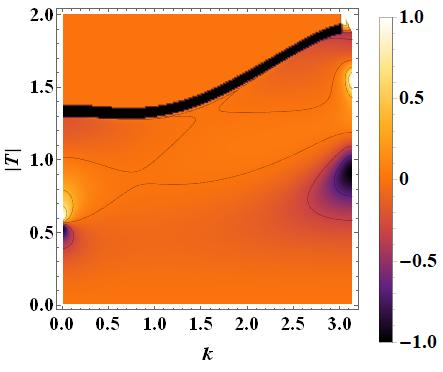}
  \end{minipage}
  \begin{minipage}[h]{0.32\linewidth}
    \centering
    \includegraphics[width=\linewidth]{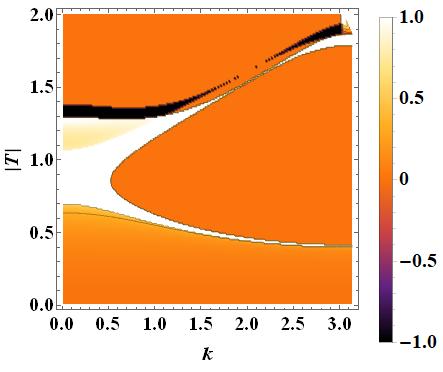}
  \end{minipage}
  \begin{minipage}[h]{0.32\linewidth}
    \centering
    \includegraphics[width=\linewidth]{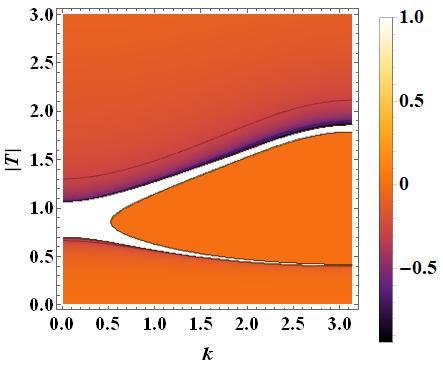}
  \end{minipage}
  \begin{minipage}[h]{0.32\linewidth}
    \centering
    \includegraphics[width=\linewidth]{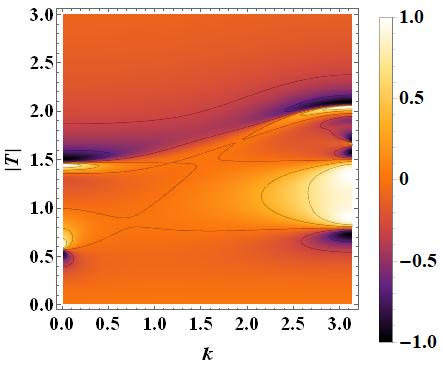}
  \end{minipage}
  \caption{Color plots of rectifying action for $\beta=0.01$ (upper row),
$\beta=0.05$ (middle row) and  $\beta=0.5$ (lower row). In two upper rows,
the three panels from left to right correspond to the first, second and third
solution, respectively. Site dependent parameter strengths $V^{(0)}=-2.5$,
$\varepsilon_V=0.05$, $\epsilon_{1}=0.5$, $\gamma_{1,2}=1$. In two upper rows,
regimes where the channel transmits only right-propagating (left-propagating)
modes are coded white (black), while regimes where the channel has no
solutions for any propagation direction are coded orange.}
 \label{Rultralow3sol}
   \end{figure}

Having identified regimes of multi-solutions, we now consider the
possibility for multi-channel asymmetric transmission
due to the presence of a small nonzero asymmetry between on-site energies.
To determine the efficiency of nonreciprocal transmission, i.e.,
transmission at diode-like modes, we define the {\em rectifying factor}
$\mathcal{R}$ as
\cite{11}
\bea{}
\mathcal{R}=\frac{t(k,|T|^2)-t(-k,|T|^2)}{t(k,|T|^2)+t(-k,|T|^2)} ,
\label{rf}
\eea
where $-1\leq\mathcal{R}\leq+1$. A perfect diode-like transmission occurs at
$\mathcal{R}=\pm1$. In Fig.\ \ref{Rultralow3sol} we
present the results of the rectifying factor along the different
transmission channels for the three different saturability regimes
(as in previous section, the results for the unsaturated case is visually
identical to those of the first two channels for the ultra-low saturation, and
thus not shown).
The
transmission scenario is presented in more detail below via plots of the
transmission coefficient $t(k,|T|^2)$ in Eq.\ \eqref{t} as a function of
transmitted {\em intensity} $|T|^2$ for the case of ultra-low saturation
($\beta=0.01$), low saturation ($\beta=0.05$) and medium saturation
($\beta=0.5$) in Fig.\ \ref{vv1}, Fig.\ \ref{k151} and Fig.\ \ref{vv1111},
respectively. Note that the blue curves in Figs.\ \ref{vv1} - \ref{vv1111}
correspond to the right-propagating case ($k>0$), while the red dotted curves
represent the left-propagation ($k<0$), both plotted in their respective
existence regimes. Note also that the first, second and third solution branches
in these figures correspond to blue, orange and green curves from
Fig.\ \ref{multisols}, respectively.


 \subsubsection*{Ultra-Low Saturation, $\beta=0.01$}

\begin{figure}[!htbp]
  \begin{minipage}[h]{0.35\linewidth}
    \centering
    \includegraphics[width=\linewidth]{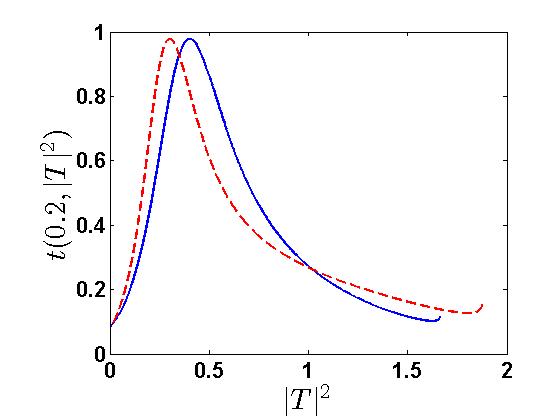}
  \end{minipage}
  \hspace{-0.7cm}
   \begin{minipage}[h]{0.35\linewidth}
    \centering
    \includegraphics[width=\linewidth]{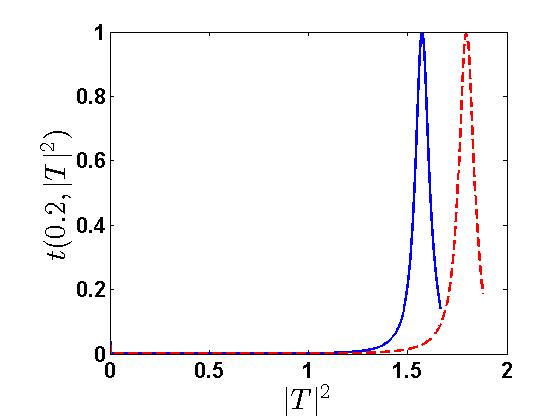}
  \end{minipage}
  \hspace{-0.7cm}
   \begin{minipage}[h]{0.35\linewidth}
    \centering
    \includegraphics[width=\linewidth]{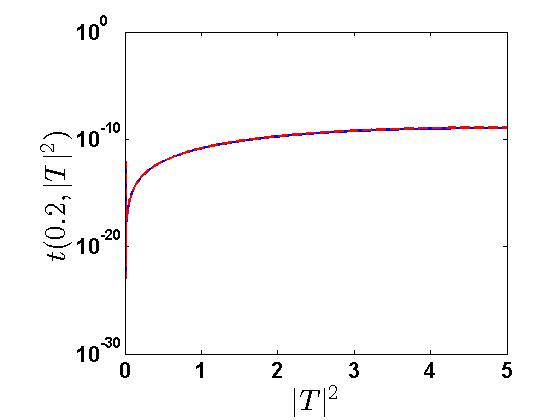}
  \end{minipage}
  \\
  \begin{minipage}[h]{0.35\linewidth}
    \centering
    \includegraphics[width=\linewidth]{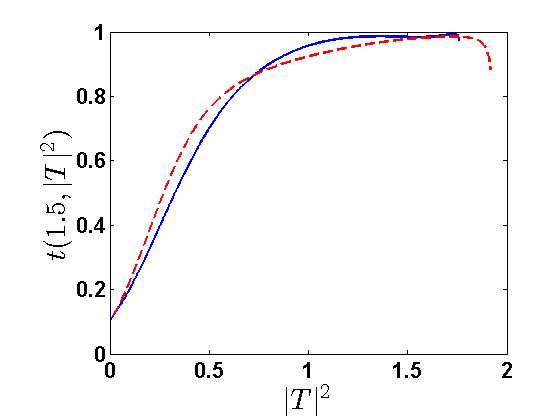}
  \end{minipage}
  \hspace{-0.7cm}
   \begin{minipage}[h]{0.35\linewidth}
    \centering
    \includegraphics[width=\linewidth]{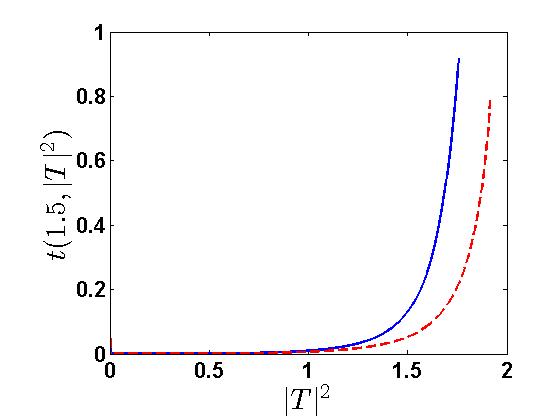}
  \end{minipage}
  \hspace{-0.7cm}
   \begin{minipage}[h]{0.35\linewidth}
    \centering
    \includegraphics[width=\linewidth]{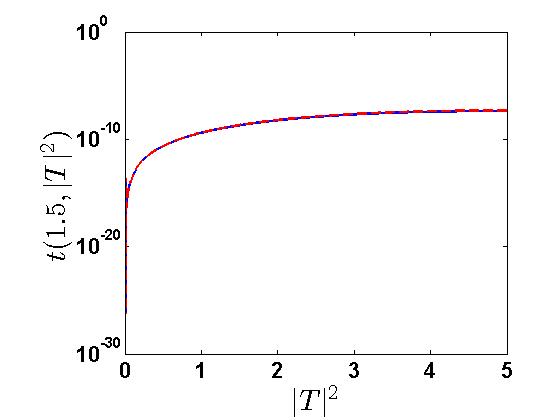}
  \end{minipage}
  \\
  \begin{minipage}[h]{0.35\linewidth}
    \centering
    \includegraphics[width=\linewidth]{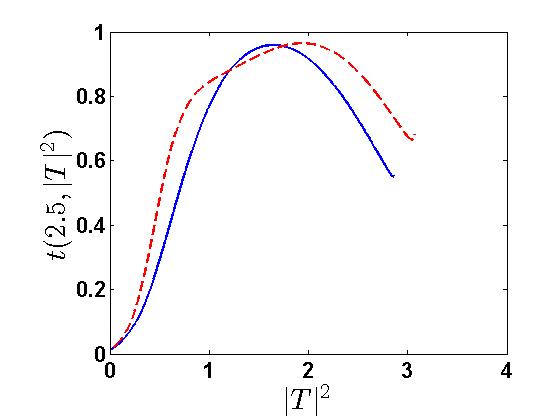}
  \end{minipage}
  \hspace{-0.7cm}
   \begin{minipage}[h]{0.35\linewidth}
    \centering
    \includegraphics[width=\linewidth]{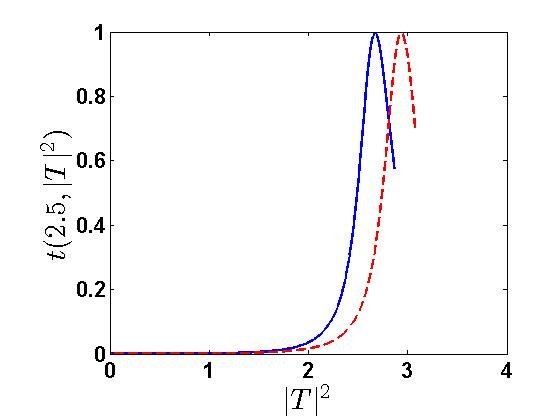}
  \end{minipage}
  \hspace{-0.7cm}
   \begin{minipage}[h]{0.35\linewidth}
    \centering
    \includegraphics[width=\linewidth]{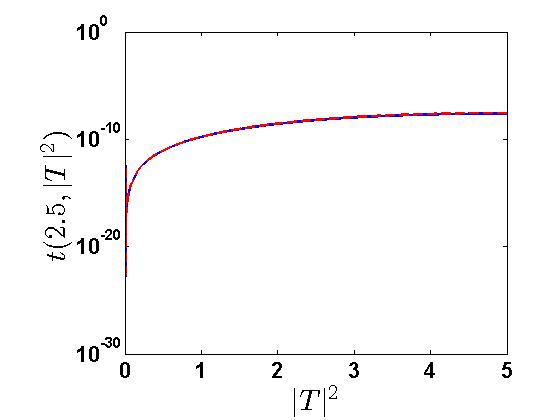}
  \end{minipage}

   \caption{Transmission coefficient $t(k,|T|^2)$ as a function of $|T|^2$
     for the ultra-low saturated case
$\beta=0.01$
     for fixed $k$ values, from top to bottom
     $k=0.2,1.5,2.5$. Left, middle and right vertical panels
  correspond to first, second and third solution branches, respectively.
     All parameter values as before.}
 \label{vv1}
  \end{figure}

For the first transmission channel, we see from upper left
Fig.\ \ref{Rultralow3sol} that there are essentially two regimes with
considerable rectification action that also correspond to large transmission
coefficients according to Fig.\ \ref{u00001}: a regime for small $k$ and
$T \approx 0.5$, and another regime for large $k$ around $T \approx 1$. As seen
in upper and lower left Fig.\ \ref{vv1}, these regimes originate in shifts
of large-amplitude transmission peaks, and are analogous to large-rectification
regimes existing for pure on-site nonlinearities \cite{11,wasay18}.
On the other hand, in the regime of intermediate $k$ where transmission peaks
are broad, transmission is close to symmetric as seen in middle left
Fig.\ \ref{vv1}. Moreover, the black band with $\mathcal{R}=-1$ in upper left
Fig.\ \ref{Rultralow3sol} for $1.3 \lesssim |T|\lesssim 1.9$ arises since
the existence regime for the first transmission channel is always
smaller for the right-propagating wave than for the left-propagating for
the corresponding set of parameter values
(see left vertical panel in  Fig.\ \ref{vv1}).

For the second transmission channel, the most interesting effect is the shift
of the narrow transmission peak with close to perfect transmission, leading to
the bright band with $\mathcal{R}$ close to 1  for
$1.1 \lesssim |T|\lesssim 1.8$ in upper middle Fig.\ \ref{Rultralow3sol}.
This is seen more
explicitly in the middle vertical panel of Fig.\ \ref{vv1}.
Note also the small regime with $\mathcal{R}$ close to -1 for $k$ close to
$\pi$ and $1.8 \lesssim |T| \lesssim 1.9$, corresponding to
the peak of almost perfect left-propagating transmission appearing
while the right-propagating transmission is declining. As the existence
regimes for the first and second solutions are identical in the regime of
ultra-low saturability, the black band with $\mathcal{R}=-1$ for the second
solution in  upper middle Fig.\ \ref{Rultralow3sol} will be the same as that
for the first solution.

Finally, as seen in upper right Fig.\ \ref{u00001} the transmission along the
third channel (corresponding to the green
solution branch in Fig.\ref{multisols}) is negligibly small, and
essentially symmetric (upper right Fig.\ \ref{Rultralow3sol} and
right vertical panel of Fig.\ \ref{vv1}).
So the system behaves as a nearly perfect mirror for this channel, in both
directions.

\subsubsection*{Low Saturation, $\beta=0.05$}

\begin{figure}[!htbp]
\begin{minipage}[h]{0.35\linewidth}
    \centering
    \includegraphics[width=\linewidth]{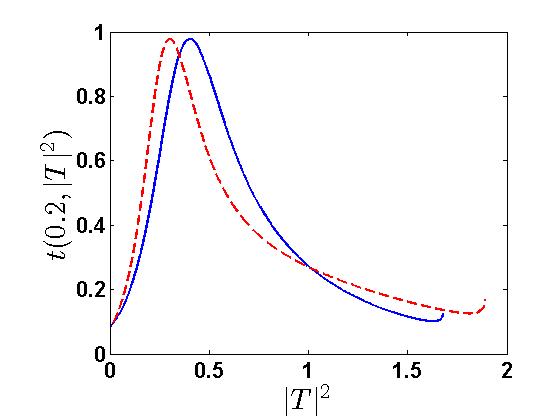}
  \end{minipage}
  \hspace{-0.7cm}
   \begin{minipage}[h]{0.35\linewidth}
    \centering
    \includegraphics[width=\linewidth]{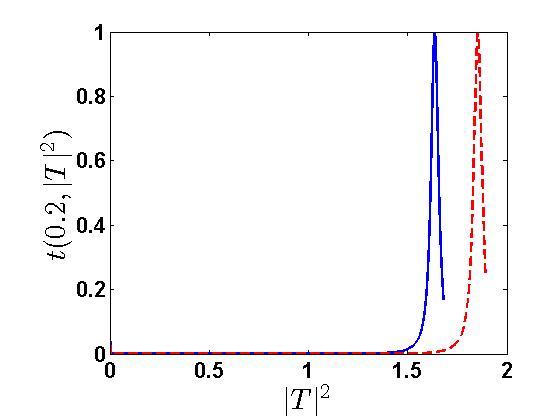}
  \end{minipage}
  \hspace{-0.7cm}
   \begin{minipage}[h]{0.35\linewidth}
    \centering
    \includegraphics[width=\linewidth]{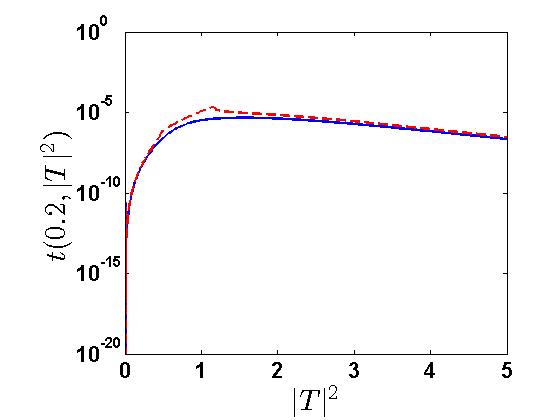}
  \end{minipage}
  \\
  \begin{minipage}[h]{0.35\linewidth}
    \centering
    \includegraphics[width=\linewidth]{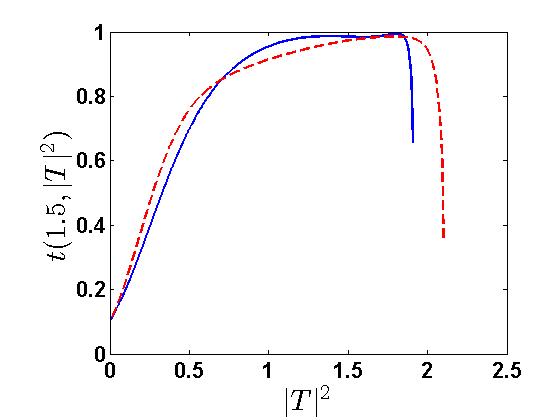}
  \end{minipage}
  \hspace{-0.7cm}
   \begin{minipage}[h]{0.35\linewidth}
    \centering
    \includegraphics[width=\linewidth]{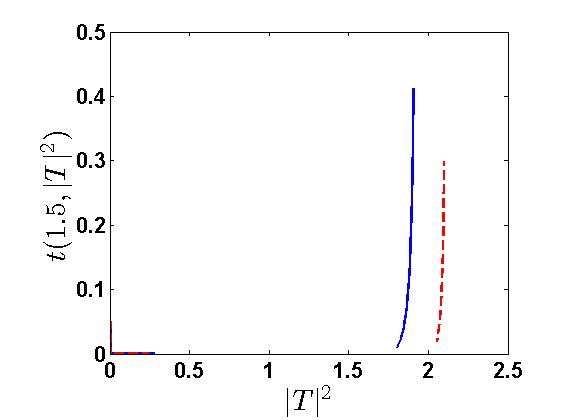}
  \end{minipage}
  \hspace{-0.7cm}
   \begin{minipage}[h]{0.35\linewidth}
    \centering
    \includegraphics[width=\linewidth]{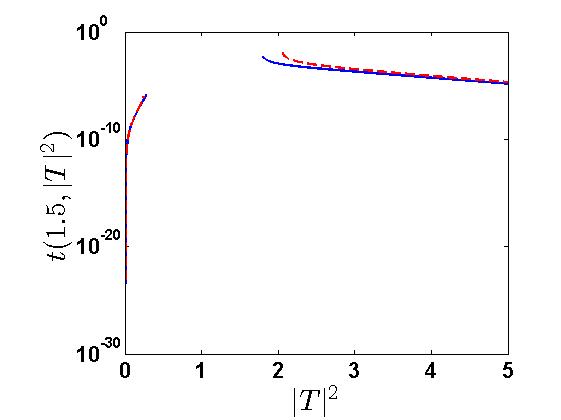}
  \end{minipage}
  \\
  \begin{minipage}[h]{0.35\linewidth}
    \centering
    \includegraphics[width=\linewidth]{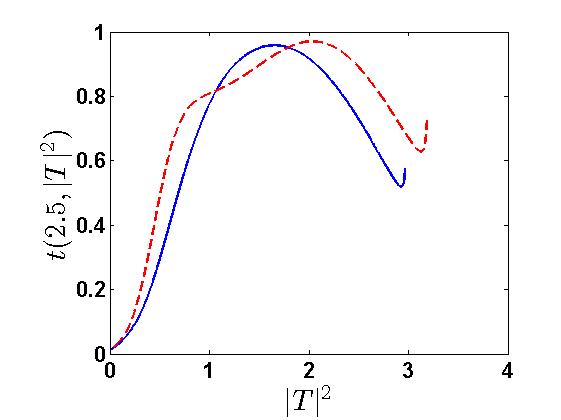}
  \end{minipage}
  \hspace{-0.7cm}
   \begin{minipage}[h]{0.35\linewidth}
    \centering
    \includegraphics[width=\linewidth]{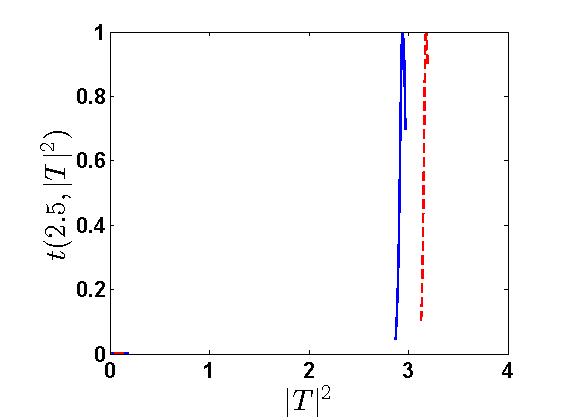}
  \end{minipage}
  \hspace{-0.7cm}
   \begin{minipage}[h]{0.35\linewidth}
    \centering
    \includegraphics[width=\linewidth]{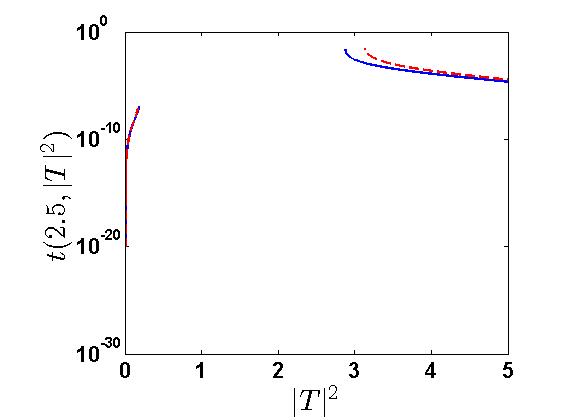}
  \end{minipage}

   \caption{
 Same as Fig. \ref{vv1} but
     for the low saturated case
     $\beta=0.05$.
All other parameters same as before.}
 \label{k151}
  \end{figure}
For the first transmission channel (middle left Fig.\ \ref{Rultralow3sol}
and left vertical panel in Fig.\ \ref{k151}), there are only minor
differences compared to the ultra-low saturation regime.

For the second channel (middle central Fig.\ \ref{Rultralow3sol}
and middle vertical panel in Fig.\ \ref{k151}), the major effects appear due to
the shrinking of its existence region, for both propagation directions. As
a result, for some $k$-values its existence regimes for left and right
propagation are fully disjoint, leading to a splitting of the  black band
in the rectification plot for $\beta=0.01$ into a ``triplet band'' with
white (only right-propagation),
orange (no propagation in any direction), and black (only left-propagation)
regions appearing in order as $|T|$ is increased. Moreover, the upper cut-off
for the low-$|T|$ regime when $k \gtrsim 0.6$ is also slightly larger for
the right-propagating wave, leading to the narrow white stripe around
$|T|\simeq 0.5$ in middle central Fig.\ \ref{Rultralow3sol}. However,
transmission is essentially negligible in both directions in the low-$|T|$
regime of the second channel (see middle vertical panel in Fig.\ \ref{k151}).

For the third channel (middle right Fig.\ \ref{Rultralow3sol}
and right vertical panel in Fig.\ \ref{k151}), regimes of non-negligible
transmission appear only close to its lower cut-off for larger $k$ (see
middle right Fig.\ \ref{u00001}), which is seen to be somewhat larger for
the left-propagating wave. The result is the upper white band in the
rectification plot corresponding to only right-propagation, followed for
slightly larger $|T|$ by a band with $\mathcal{R}$ close to -1. The lower
white stripe is the same as for the second solution since their
small-$|T|$ existence regimes are identical.

\subsubsection*{Medium Saturation, $\beta=0.5$}

\begin{figure}[!htbp]
  \begin{minipage}[h]{0.35\linewidth}
    \centering
    \includegraphics[width=\linewidth]{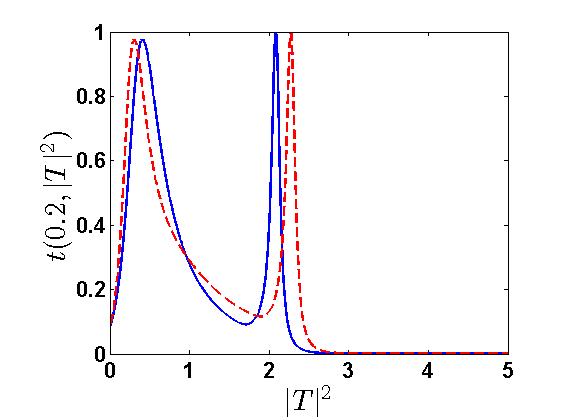}
  \end{minipage}
  \hspace{-0.7cm}
  \begin{minipage}[h]{0.35\linewidth}
    \centering
    \includegraphics[width=\linewidth]{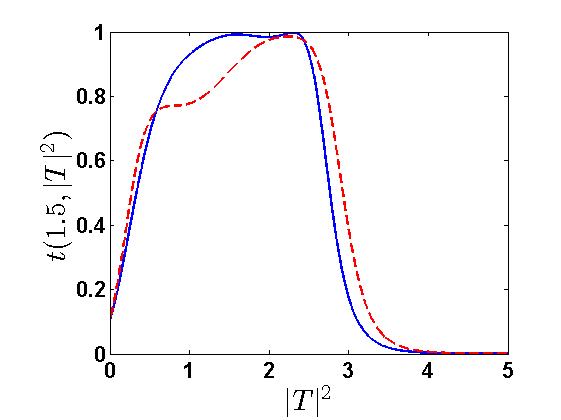}
  \end{minipage}
  \hspace{-0.7cm}
  \begin{minipage}[h]{0.35\linewidth}
    \centering
    \includegraphics[width=\linewidth]{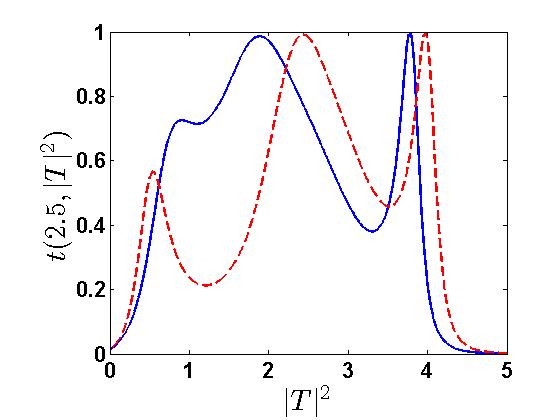}
  \end{minipage}

  \caption{Transmission coefficient $t(k,|T|^2)$ as a function of $|T|^2$
    for the medium saturated case $\beta=0.5$,  for fixed $k$ values
    $k=0.2$ (left), $k=1.5$ (middle), $k=2.5$ (right)}.
 \label{vv1111}
  \end{figure}
In this regime, the multi-solution regime is entirely suppressed and thus
there is only a single solution for all $k$ and $|T|$. We
present the rectifying action in lower Fig.\ \ref{Rultralow3sol} and
corresponding plots of transmission coefficient in Fig.\ \ref{vv1111}.
The plots are qualitatively similar to those for pure on-site
nonlinearity \cite{11,wasay18}. Note that the previous black bands in the
rectification plots for
the first and second solutions in regimes of smaller saturability now have
turned into a dark band with $\mathcal{R}$ close to -1, as the narrow
transmission peaks in the high-$|T|$ regime
now appear for the same solution branch (left Fig.\ \ref{vv1111}). The
additional transmission peaks appearing for large $k$
(lower Fig.\ \ref{u00001} and right Fig.\ \ref{vv1111}) also lead to
a more complicated pattern of regimes with $\mathcal {R}$ alternating
between values close to +1 and -1 for increasing $|T|$, when $k$ is close to
$\pi$.

\section{Stability Analysis}
\label{sec:stability}

In this section, we investigate the dynamical stability of the stationary
solutions used in previous sections. The stability analysis will closely
follow as done for instance in \cite{lepri,jd} for pure on-site
nonlinearities, and it is performed by first perturbing the solutions as
\bea{}
A_n(t)=P_n(t)+\delta R_n(t) .
\label{perturbation}
\eea
This will linearize the equation of motion \eqref{dynamical}. The resulting
linearized set of dynamical equations of order $\delta$ is
\bea{}
i\dot{R_n}-V_nR_n+R_{n+1}+R_{n-1}
=
\gamma_n\left(2R_n\!\mid\!P_n\!\mid^2\!+~P_n^2R_n^\ast \right)+~~~~~~~~~~~~~\nonumber
\\
\epsilon_{n}\left[\frac{W_{n+1} P_n}{(1+\beta|P_{n+1}|^2|P_{n}|^2)^2}-\frac{|P_{n+1}|^4 P_n\beta Z}{(1+\beta|P_{n+1}|^2|P_{n}|^2)^2}+\frac{R_n|P_{n+1}|^2}{1+\beta|P_{n+1}|^2|P_{n}|^2}\right]+~~~~~~\nonumber
\\
\epsilon_{n-1}\left[\frac{W_{n-1} P_n}{(1+\beta|P_{n-1}|^2|P_{n}|^2)^2}-\frac{|P_{n-1}|^4 P_n\beta Z}{(1+\beta|P_{n-1}|^2|P_{n}|^2)^2}+\frac{R_n|P_{n-1}|^2}{1+\beta|P_{n-1}|^2|P_{n}|^2} \right] ,
\label{linearized}
\eea
where $W_{n+1}=P^\ast_{n+1}R_{n+1}+R^\ast_{n+1}P_{n+1}$, $W_{n-1}=P^\ast_{n-1}R_{n-1}+R^\ast_{n-1}P_{n-1}$ and $Z=P^\ast_nR_n+R^\ast_nP_n$.

With the ansatz $P_n(t)=A_ne^{-i\omega t},~A_n\neq A_n(t)$ being the
complex amplitudes of a stationary solution as before,
and $R_n(t)=e^{-i\omega t}(a_n e^{i\nu t}+b_n e^{-i\nu^\ast t})$,
the linearized set of equations \eqref{linearized} yields an eigenvalue
problem of the following form
\begin{gather}
 \nu\begin{bmatrix} a_n \\ b^\ast_n \end{bmatrix}
 =
   \begin{bmatrix}
   M_1  &
   M_2 \\
   M_3 &
   M_4
   \end{bmatrix}.\begin{bmatrix} a_n \\ b^\ast_n \end{bmatrix} .
   \label{evals}
\end{gather}
Specifying to the two nonlinear sites (dimer), the resulting matrices are
\begin{gather}
\hspace{-1cm} M_1
 =
   \begin{bmatrix}
   \omega-V_1-2\gamma_1|A_1|^2+\frac{\epsilon_1|A_2|^4|A_1|^2\beta}{(1+\beta|A_1|^2|A_2|^2)^2}-\frac{\epsilon_1|A_2|^2}{1+\beta|A_2|^2|A_1|^2}  &
   1-\frac{\epsilon_1A_2^\ast A_1}{(1+\beta|A_2|^2|A_1|^2)^2} \\
   1-\frac{\epsilon_1A_1^\ast A_2}{(1+\beta|A_1|^2|A_2|^2)^2} &
   \omega-V_2-2\gamma_2|A_2|^2+\frac{\epsilon_1|A_1|^4|A_2|^2\beta}{(1+\beta|A_1|^2|A_2|^2)^2}-\frac{\epsilon_1|A_1|^2}{1+\beta|A_1|^2|A_2|^2}
   \end{bmatrix}\nonumber
\end{gather}
\begin{gather}
 M_2
 =
   \begin{bmatrix}
   -\gamma_1 A_1^2+\frac{\epsilon_1|A_2|^4 A_1^2\beta}{(1+\beta|A_1|^2|A_2|^2)^2}  &
   \frac{-\epsilon_1A_2 A_1}{(1+\beta|A_2|^2|A_1|^2)^2} \\
   \frac{-\epsilon_1A_1 A_2}{(1+\beta|A_1|^2|A_2|^2)^2} &
   -\gamma_2 A_2^2+\frac{\epsilon_1|A_1|^4 A_2^2\beta}{(1+\beta|A_1|^2|A_2|^2)^2}
   \end{bmatrix}\nonumber
\end{gather}
\begin{gather}
 M_3
 =
   \begin{bmatrix}
   \gamma_1 A_1^{\ast 2}-\frac{\epsilon_1|A_2|^4 A_1^{\ast 2}\beta}{(1+\beta|A_1|^2|A_2|^2)^2}  &
   \frac{\epsilon_1A^\ast_2 A^\ast_1}{(1+\beta|A_2|^2|A_1|^2)^2} \\
   \frac{\epsilon_1A^\ast_1 A^\ast_2}{(1+\beta|A_1|^2|A_2|^2)^2} &
   \gamma_2 A_2^{\ast 2}-\frac{\epsilon_1|A_1|^4 A_2^{\ast 2}\beta}{(1+\beta|A_1|^2|A_2|^2)^2}
   \end{bmatrix}\nonumber
\end{gather}
and
\begin{gather}
\hspace{-1cm} M_4
 \!=\!
   \begin{bmatrix}
   -\omega\!+\!V^\ast_1\!+2\gamma_1|A_1|^2-\frac{\epsilon_1|A_2|^4|A_1|^2\beta}{(1+\beta|A_1|^2|A_2|^2)^2}+\frac{\epsilon_1|A_2|^2}{1+\beta|A_2|^2|A_1|^2}  &
   -1+\frac{\epsilon_1A_2 A^\ast_1}{(1+\beta|A_2|^2|A_1|^2)^2} \\
   -1+\frac{\epsilon_1A_1 A^\ast_2}{(1+\beta|A_1|^2|A_2|^2)^2} &
   -\omega\!+\!V^\ast_2\!+2\gamma_2|A_2|^2-\frac{\epsilon_1|A_1|^4|A_2|^2\beta}{(1+\beta|A_1|^2|A_2|^2)^2}+\frac{\epsilon_1|A_1|^2}{1+\beta|A_1|^2|A_2|^2}
   \end{bmatrix} .
\nonumber
\end{gather}

 For lattices of size $m$ (with $m=2n+1$ where $n$ is the site counter for
each linear side) with this type of nonlinear dimer embedded inside the linear
chains to its right and left, the matrices ($M_{1,..,4}$) will be higher
dimensional, i.e., $m\times m$. However, apart from
 the dimer sites the site-dependent coefficients are zero, therefore, in the
linear region all entries in $M_2$ and $M_3$ will be zero, and
 \bea{}
 M_1&=&diag(\omega)+G \nonumber
 \\
 M_4&=&diag(-\omega)-G\nonumber ,
 \eea
 where $G$ is an $m\times m$ sparse matrix with ones on both super- and
sub-diagonals. The eigenvalue problem corresponding to Eq.\ \eqref{evals} is
then essentially the eigenvalue computation of the resulting $2m\times 2m$
sparse matrix. This is done by direct numerical computation of the eigenvalues
for a finite-sized chain with the nonlinear dimer embedded in the center.

 We will present below sample results of this stability analysis for all
multi-solution saturation regimes. The eigenvalue/eigenvector computations
will be shown for a lattice of 201 sites, and the corresponding time
propagation computations have been performed for lattices of up to 2001 sites,
in
order to accommodate for large $t$-values and avoid boundary errors. As pointed
out in \cite{lepri}, extended eigenvectors corresponding to a continuous
spectrum in the infinite-chain limit may cause spurious instabilities due to
boundary effects, of the order $1/m$ ($\sim 5\times 10^{-3}$ below). Solutions
exhibiting only such unstable eigenmodes thus correspond to linearly
stable scattering solutions for the original set-up.

 \subsubsection*{Ultra Low Saturation, $\beta=0.01$}

\begin{figure}[!htbp]
\begin{minipage}[h]{0.32\linewidth}
    \centering
    \includegraphics[width=\linewidth]{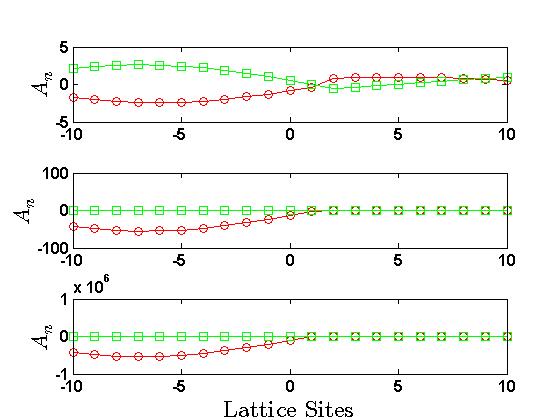}
  \end{minipage}
  \begin{minipage}[h]{0.32\linewidth}
    \centering
    \includegraphics[width=\linewidth]{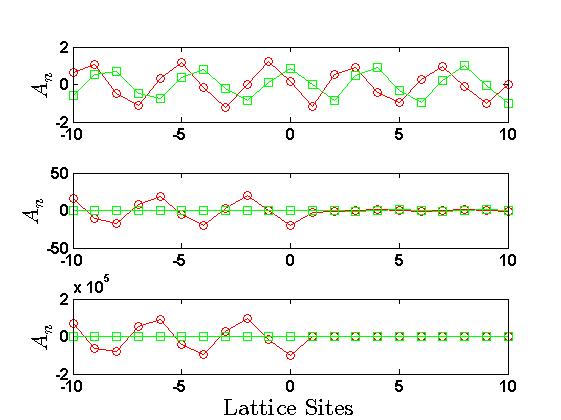}
  \end{minipage}
  \begin{minipage}[h]{0.32\linewidth}
    \centering
    \includegraphics[width=\linewidth]{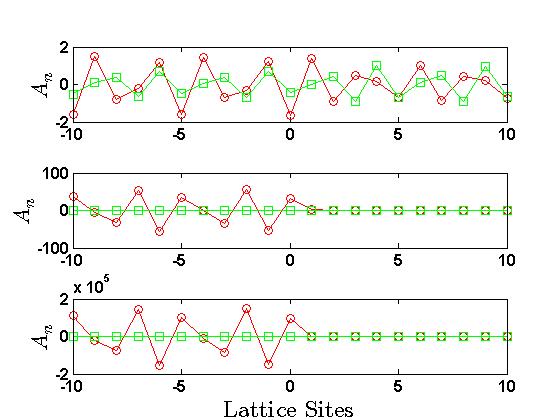}
  \end{minipage}

   \caption{Stationary solutions at $\beta=0.01$, $|T|=1$, real (red circles)
and imaginary (green squares) part of the solutions: left vertical panel for
$k=0.2$, middle vertical panel for $k=1.5$ and right vertical panel for
$k=2.5$. In each vertical panel,
top plot corresponds to the first solution branch, middle to the second
branch and lower to the third branch.}
 \label{ssb01}
  \end{figure}

We have chosen $|T|=1$ and $k=0.2,1.5,2.5$ to represent solutions
corresponding to small, medium and large wavenumbers respectively, in order to
connect with our earlier discussion.
The stationary solutions
are depicted in Fig.\ \ref{ssb01}, for the three solution branches.
%
Note that in this regime, the transmission coefficient is considerable only
for the first solution, very small for the second and essentially negligible
for the third, for all values of $k$.
%
%
%

\begin{figure}[!htbp]
  \begin{minipage}[h]{0.32\linewidth}
    \centering
    \includegraphics[width=\linewidth]{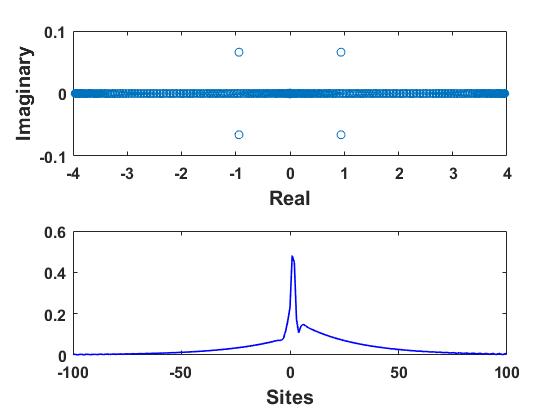}
  \end{minipage}
  \begin{minipage}[h]{0.32\linewidth}
    \centering
    \includegraphics[width=\linewidth]{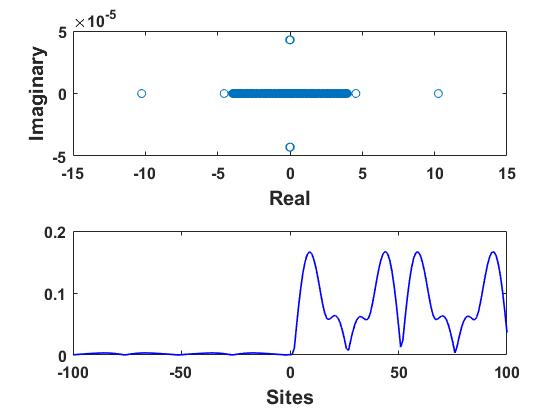}
  \end{minipage}
  \begin{minipage}[h]{0.32\linewidth}
    \centering
    \includegraphics[width=\linewidth]{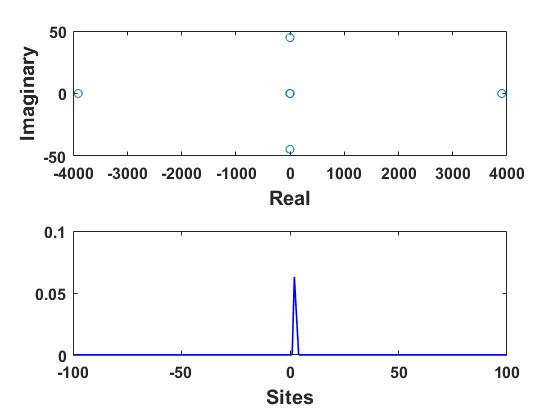}
  \end{minipage}
  \begin{minipage}[h]{0.32\linewidth}
    \centering
    \includegraphics[width=\linewidth]{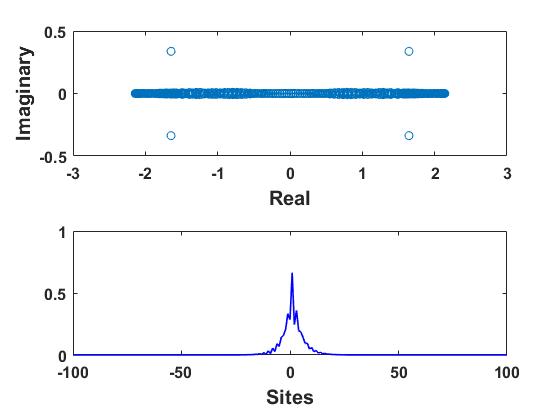}
  \end{minipage}
  \begin{minipage}[h]{0.32\linewidth}
    \centering
    \includegraphics[width=\linewidth]{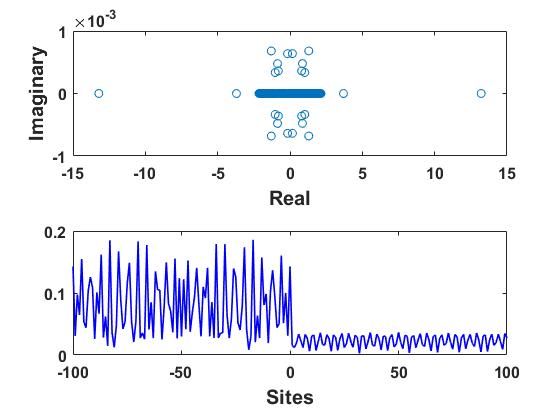}
  \end{minipage}
  \begin{minipage}[h]{0.32\linewidth}
    \centering
    \includegraphics[width=\linewidth]{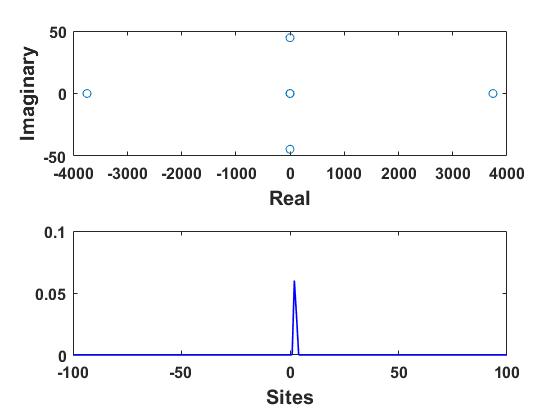}
  \end{minipage}
\\
  \begin{minipage}[h]{0.32\linewidth}
    \centering
    \includegraphics[width=\linewidth]{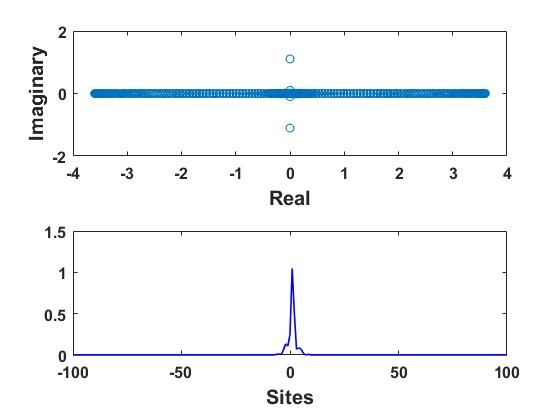}
  \end{minipage}
  \begin{minipage}[h]{0.32\linewidth}
    \centering
    \includegraphics[width=\linewidth]{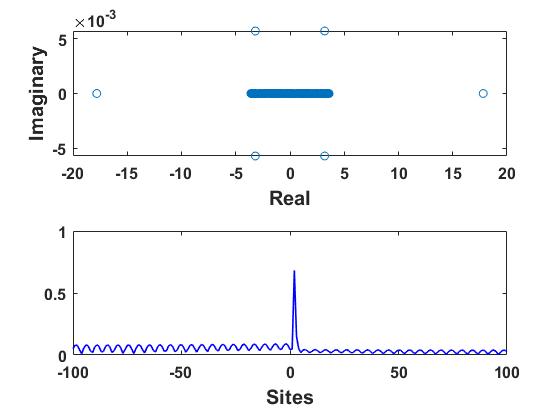}
  \end{minipage}
  \begin{minipage}[h]{0.32\linewidth}
    \centering
    \includegraphics[width=\linewidth]{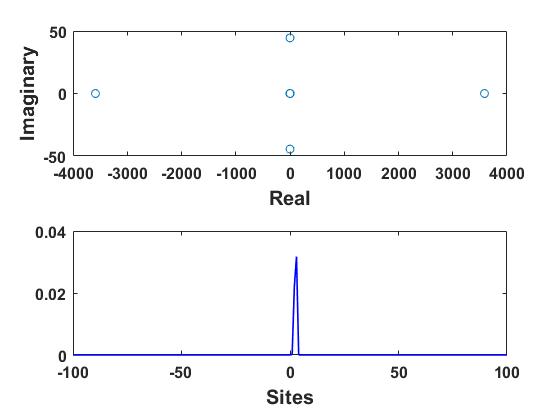}
  \end{minipage}
     \caption{Stability analysis for $|T|=1$ at $\beta=0.01$:
$k=0.2$ (first and second row), $k=1.5$ (third and fourth row),
$k=2.5$ (fifth and sixth row).
Left vertical panel shows the eigenvalues and most unstable
eigenvectors for first solution,
middle panel for the second solution and right panel for the third solution.}
 \label{b01evals}
  \end{figure}

Results from the stability analysis
are presented in Fig.\ \ref{b01evals}.
For the first solution (left vertical panel), we find that it is always
unstable for these parameter values, with an unstable eigenvector localized
at the dimer. For small and intermediate $k$, the unstable
eigenvalues are complex
corresponding to a rather weak oscillatory instability, while a stronger
purely exponential instability appears for larger $k$. The second solution
(middle vertical panel)
is essentially stable for small and intermediate $k$ (the observed tiny
imaginary parts of eigenvalues correspond to extended eigenvectors and are
spurious due to boundary effects as discussed above), while a very weak,
oscillatory instability appears for larger $k$ due to a resonance between
a mode localized at the dimer and the continuous spectrum. The third solution
(right vertical panel) is strongly unstable with a purely imaginary
eigenvalue, and an eigenmode strongly localized at site 2.

\begin{figure}[!htbp]
  \begin{minipage}[h]{0.32\linewidth}
    \centering
    \includegraphics[width=\linewidth]{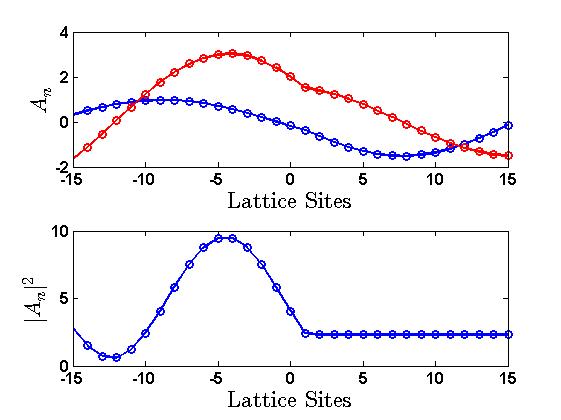}
  \end{minipage}
   \begin{minipage}[h]{0.32\linewidth}
    \centering
   \includegraphics[width=\linewidth]{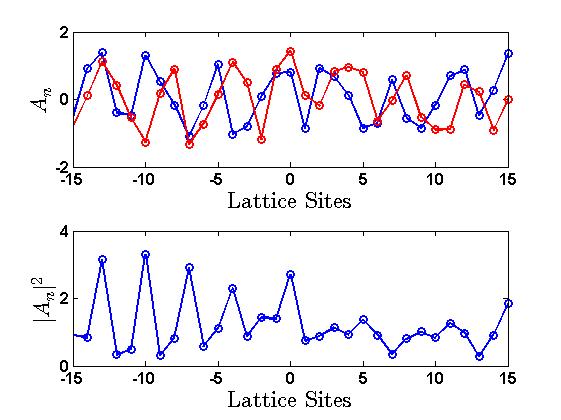}
  \end{minipage}
   \begin{minipage}[h]{0.32\linewidth}
    \centering
    \includegraphics[width=\linewidth]{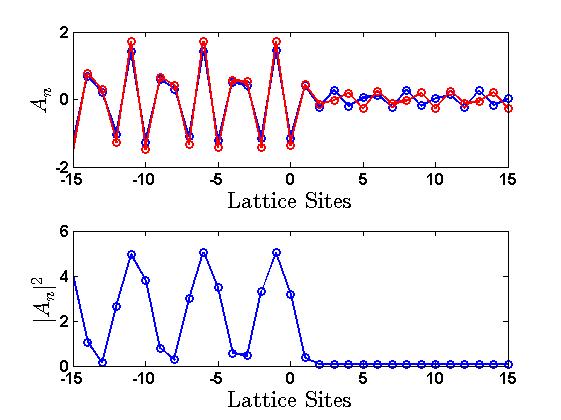}
  \end{minipage}
   \caption{Time propagation corresponding to the first solution
in Figs.\ \ref{ssb01}-\ref{b01evals};
$k=0.2$, $k=1.5$, $k=2.5$ from left to right.
Snapshot at time 400 with a small $\mathcal{O}(10^{-3})$ arbitrary perturbation
inserted at site 1.
Upper figures: Real (blue) and imaginary (red) parts. Lower: $|A_n|^2$.
}
 \label{tpropb01sol1}
  \end{figure}

The dynamics resulting from the instabilities of the first solution is
illustrated in Fig.\ref{tpropb01sol1}.
%
For $k=0.2$ we observe that, after an initially oscillatory dynamics with
exponentially increasing amplitudes at the center, the solution settles down
into an almost stationary transmission regime corresponding to a
{\em larger} $|T|$ ($|A_2|\approx 1.52$ at time 400 in left
Fig.\ \ref{tpropb01sol1}). However, this value of
$|T|$ is above the existence threshold for the first solution at $k=0.2$, and
the solution is not strictly stationary but shows small-amplitude oscillations
for $|A|$ at the dimer sites, as well as a weak long-wavelength spatial
modulation. At $k=1.5$ (middle Fig.\ \ref{tpropb01sol1}) the oscillatory
instability is stronger and yields persistent spatiotemporal oscillations,
while at  $k=2.5$ (right Fig.\ \ref{tpropb01sol1}) the solution instead, after
the initial strong non-oscillatory exponential instability, settles down into
an essentially stationary state with considerably {\em smaller} $|T|$
($|A_2|\approx 0.27$ at time 400).

The second solution is essentially stable for $|T|=1$, and no significant
changes are observed in the time evolution for any of the considered values
of $k$.
\begin{figure}[!htbp]
  \begin{minipage}[h]{0.32\linewidth}
    \centering
    \includegraphics[width=\linewidth]{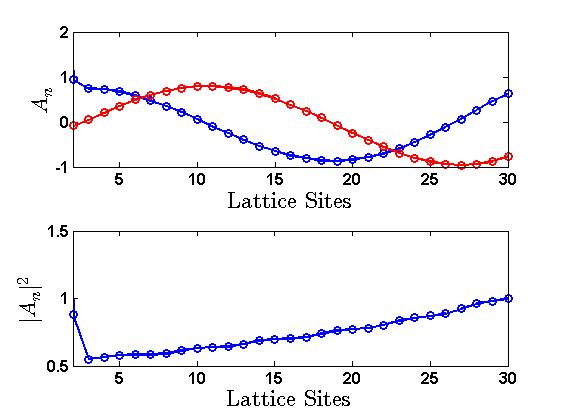}
  \end{minipage}
   \begin{minipage}[h]{0.32\linewidth}
    \centering
   \includegraphics[width=\linewidth]{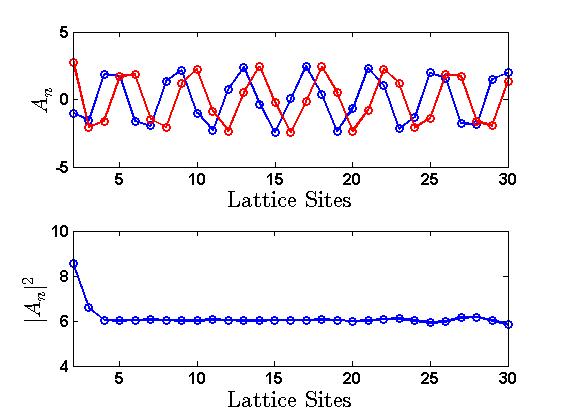}
  \end{minipage}
   \begin{minipage}[h]{0.32\linewidth}
    \centering
    \includegraphics[width=\linewidth]{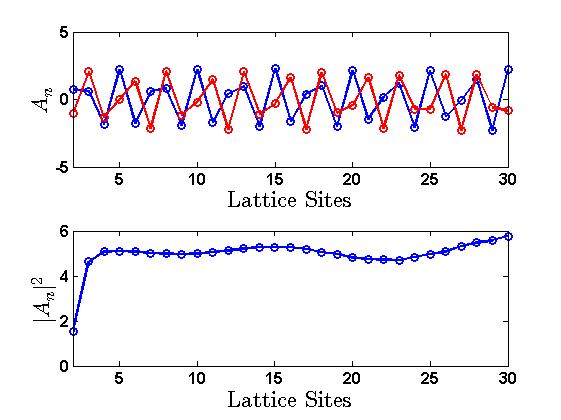}
  \end{minipage}
   \caption{
     Similar as  Fig.\ \ref{tpropb01sol1} but for
       the third solution
in Figs.\ \ref{ssb01}-\ref{b01evals},
with snapshots at time 40.
Only sites 2-30 of a 401-site system are shown.
}
 \label{tpropb01sol3}
  \end{figure}
For the third solution, instability-induced dynamics is
illustrated in Fig.\ \ref{tpropb01sol3}, where the amplitudes
of the initial portion of the transmitting side only (amplitudes at the
incoming side are of the order of $10^5$ as in Fig.\ \ref{ssb01}) are
shown at time 40
(the instability develops rapidly due to the large eigenvalue).
Typically,
the instability results in an initial decrease of $|A_2|$ (where the unstable
eigenmode is localized) to values close to zero, followed by recurring
oscillations between small and larger amplitudes at this site. As a result of
these oscillations the transmitted intensity will start to deviate from 1 as
seen previously for the first solution branch; note however from
Fig.\ \ref{tpropb01sol3} that now $|A_n|<1$ for $k=0.2$ and $|A_n|>1$ for
$k=1.5$ and $k=2.5$ at the initial portion of the transmitting side
(the transmission coefficient evidently remains very small
due to the huge amplitudes on the left side).






  \subsubsection*{Low saturation, $\beta=0.05$}

\begin{figure}[!htbp]
  \begin{minipage}[h]{0.32\linewidth}
    \centering
    \includegraphics[width=\linewidth]{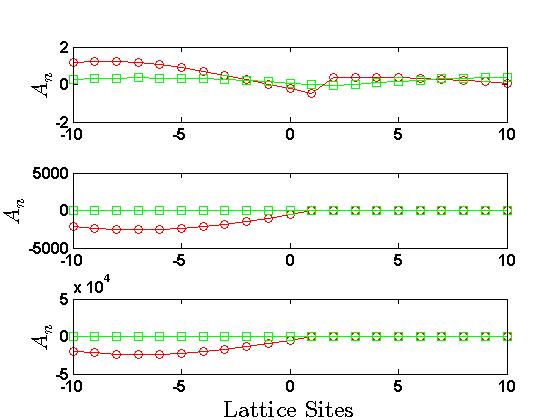}
  \end{minipage}
  \begin{minipage}[h]{0.32\linewidth}
    \centering
    \includegraphics[width=\linewidth]{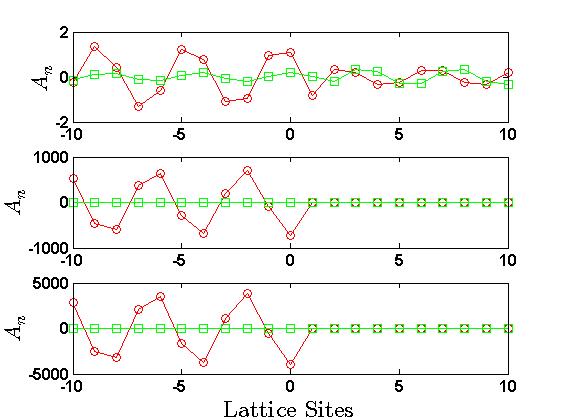}
  \end{minipage}
  \begin{minipage}[h]{0.32\linewidth}
    \centering
    \includegraphics[width=\linewidth]{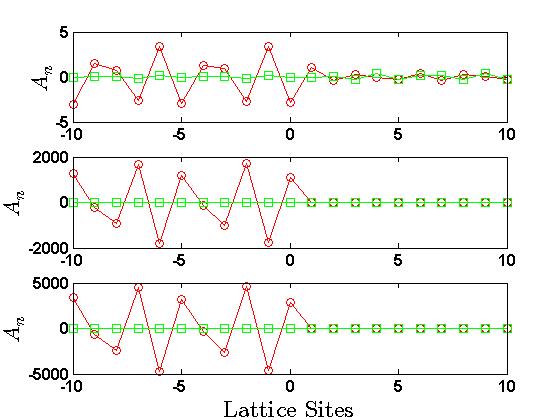}
  \end{minipage}
  \caption{
Same as Fig.\ \ref{ssb01} but with $\beta=0.05$ and $|T|=0.4$.
  }
 \label{ssb05}
  \end{figure}

We will here pick $|T|=0.4$ as representative for the small-$T$ regime giving
three different solutions for each of the sample $k$ values
($k=0.2, 1.5, 2.5$),
and show the stationary solution plots in Fig.\ \ref{ssb05}.
Note that in this regime, the transmission coefficient is non-negligible only
for the first solution.

\begin{figure}[!htbp]
  \begin{minipage}[h]{0.32\linewidth}
    \centering
    \includegraphics[width=\linewidth]{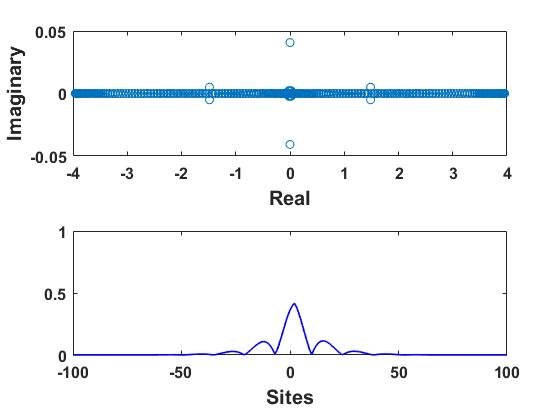}
  \end{minipage}
  \begin{minipage}[h]{0.32\linewidth}
    \centering
    \includegraphics[width=\linewidth]{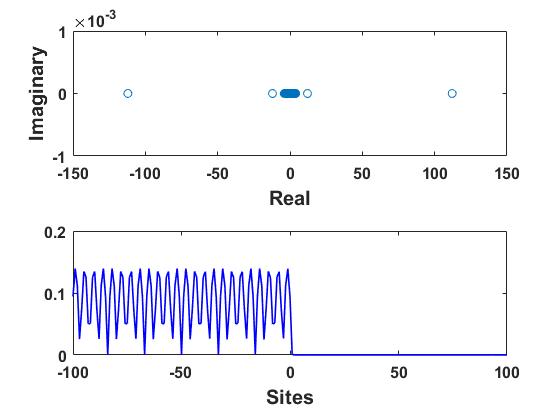}
  \end{minipage}
  \begin{minipage}[h]{0.32\linewidth}
    \centering
    \includegraphics[width=\linewidth]{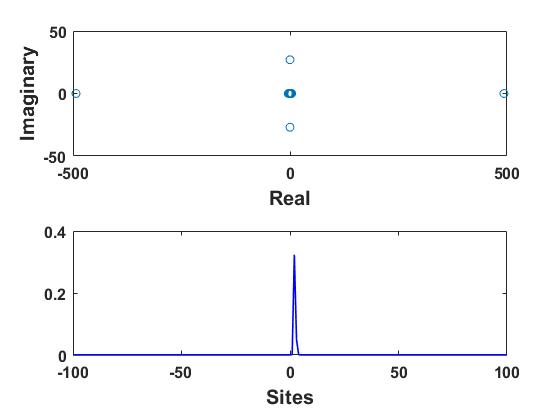}
  \end{minipage}
  \\
  \begin{minipage}[h]{0.32\linewidth}
    \centering
    \includegraphics[width=\linewidth]{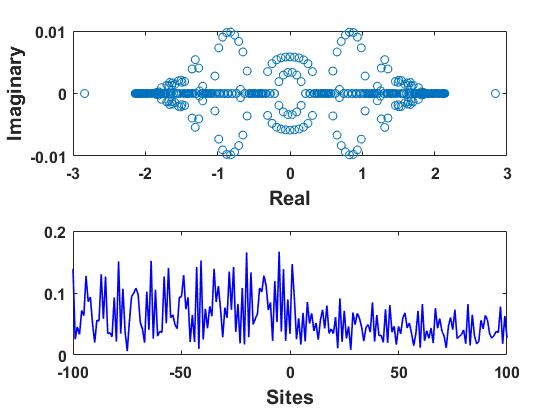}
  \end{minipage}
  \begin{minipage}[h]{0.32\linewidth}
    \centering
    \includegraphics[width=\linewidth]{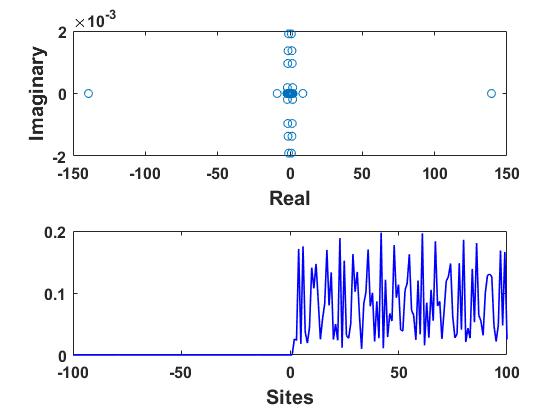}
  \end{minipage}
  \begin{minipage}[h]{0.32\linewidth}
    \centering
    \includegraphics[width=\linewidth]{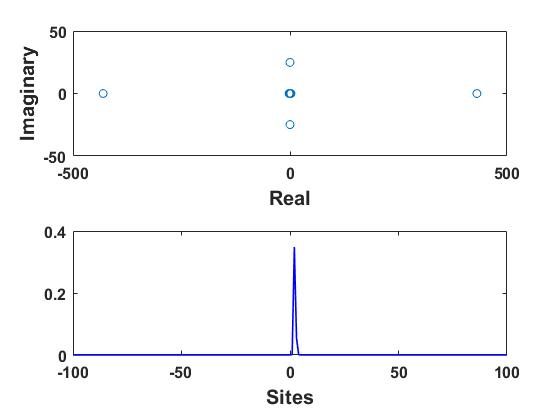}
  \end{minipage}
\\
  \begin{minipage}[h]{0.32\linewidth}
    \centering
    \includegraphics[width=\linewidth]{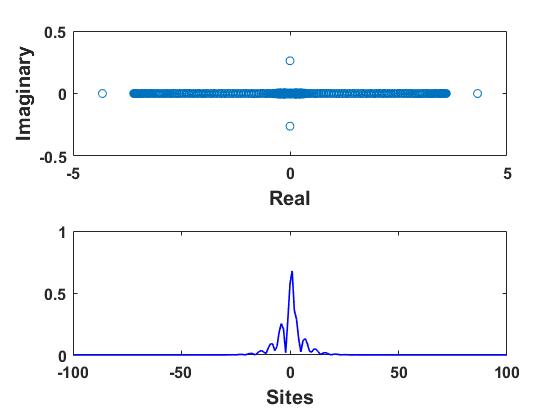}
  \end{minipage}
  \begin{minipage}[h]{0.32\linewidth}
    \centering
    \includegraphics[width=\linewidth]{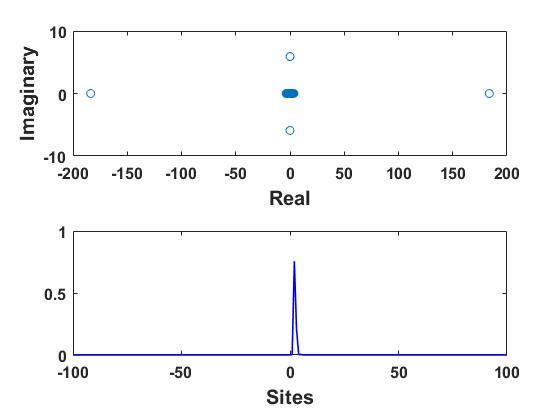}
  \end{minipage}
  \begin{minipage}[h]{0.32\linewidth}
    \centering
    \includegraphics[width=\linewidth]{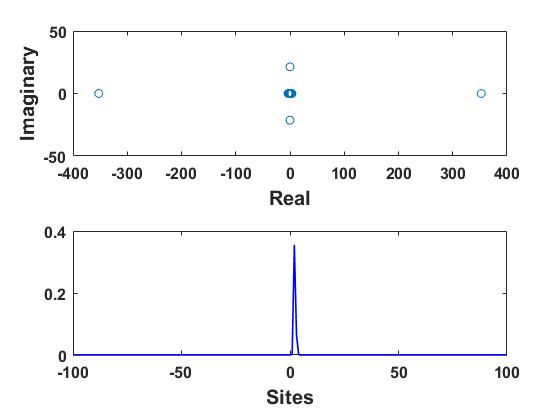}
  \end{minipage}
  \\

  \caption{
    Same as Fig.\ \ref{b01evals} but with $\beta=0.05$ and
      $|T|=0.4$.
}
 \label{b05evals}
  \end{figure}

The stability of each solution in this regime at the respective $k$ values is
illustrated in Fig.\ \ref{b05evals}.
Comparing to Fig.\ \ref{b01evals}, we note that the instabilities for the first
solution here are generally weaker, and to our numerical accuracy it is
linearly stable at $k=1.5$ (i.e., for wave numbers close to $\pi/2$). The
second solution is still stable for small and intermediate wave numbers, but
now destabilizes with a purely imaginary eigenvalue for larger $k$, where it
is close to its bifurcation point with the third solution
(see Fig.\ \ref{multisols}). The third solution remains strongly unstable.

\begin{figure}[!htbp]
  \begin{minipage}[h]{0.32\linewidth}
    \centering
    \includegraphics[width=\linewidth]{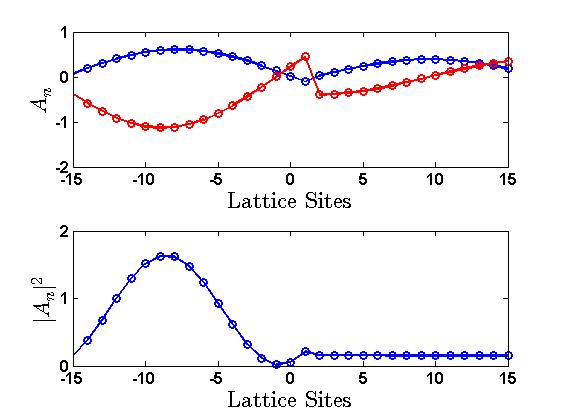}
  \end{minipage}
  \begin{minipage}[h]{0.32\linewidth}
    \centering
    \includegraphics[width=\linewidth]{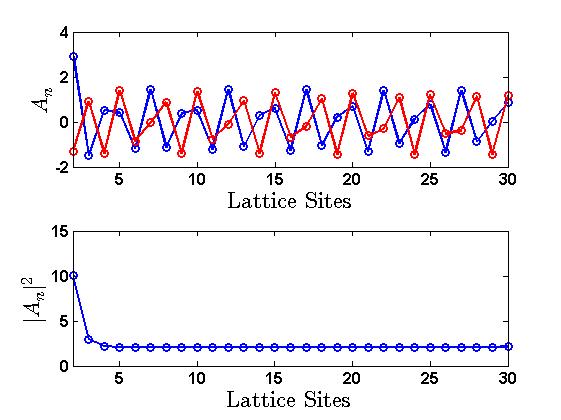}
  \end{minipage}
   \caption{Time propagation of unstable stationary solutions
from Figs.\ \ref{ssb05}-\ref{b05evals};
Left: First solution, $k=0.2$.
Right: Second solution, $k=2.5$ (only sites 2-30 are shown).
Snapshots at time 400
with a small $\mathcal{O}(10^{-3})$ arbitrary perturbation
inserted at site 1. Upper figures: Real (blue) and imaginary (red) parts.
Lower: $|A_n|^2$.
}
 \label{tpropb05}
  \end{figure}

To illustrate the outcome of these instabilities, we show in Fig.\
\ref{tpropb05} snapshots
of solutions only for unstable cases that differ significantly to those
previously shown for $\beta=0.01$ and $|T|=1$. For the first solution at
$k=0.2$ (left Fig.\ \ref{tpropb05})
the instability is now non-oscillatory, and results in a slight
decrease of the amplitude on the transmitting side
($|A_2|\approx 0.39$ at time 400). At $k=1.5$ the solution is stable, and
at $k=2.5$ the unstable dynamics is analogous to that of
Fig.\ \ref{tpropb01sol1}. The unstable dynamics of the second solution
at $k=2.5$ is illustrated in right Fig.\ \ref{tpropb05} and is similar to
that previously described for the third solution, resulting after some
transient in amplitudes $|A_n|>0.4$ at the initial portion of the transmitting
side (the transmission coefficient remaining small with maximal amplitudes
 $|A_n|\sim 2000$ on the left side).
For the third solution, dynamics is analogous
to that observed in Fig.\ \ref{tpropb01sol3}.


  \subsubsection*{Medium Saturation, $\beta=0.5$}

\begin{figure}[!htbp]
  \begin{minipage}[h]{0.32\linewidth}
    \centering
    \includegraphics[width=\linewidth]{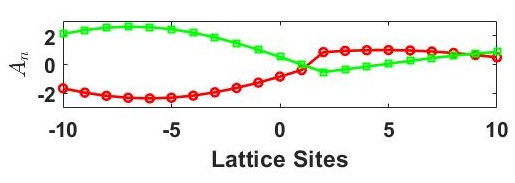}
  \end{minipage}
  \begin{minipage}[h]{0.32\linewidth}
    \centering
    \includegraphics[width=\linewidth]{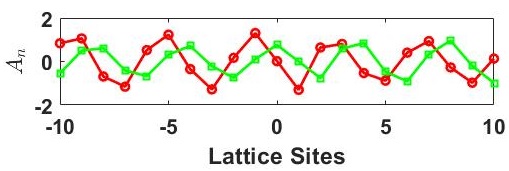}
  \end{minipage}
  \begin{minipage}[h]{0.32\linewidth}
    \centering
    \includegraphics[width=\linewidth]{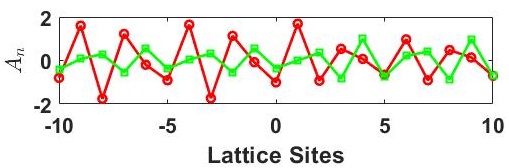}
  \end{minipage}
  \\
  \begin{minipage}[h]{0.32\linewidth}
    \centering
    \includegraphics[width=\linewidth]{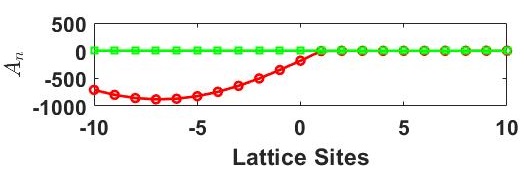}
  \end{minipage}
  \begin{minipage}[h]{0.32\linewidth}
    \centering
    \includegraphics[width=\linewidth]{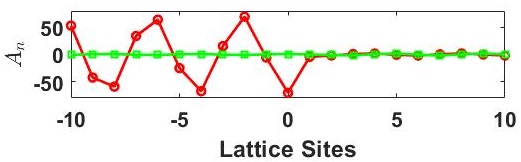}
  \end{minipage}
  \begin{minipage}[h]{0.32\linewidth}
    \centering
    \includegraphics[width=\linewidth]{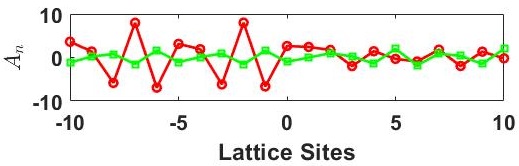}
  \end{minipage}
  \caption{Stationary solutions at $\beta=0.5$, real (red circles) and
imaginary (green squares) part of the solutions: Left, middle and right
vertical panels for $k=0.2$, $k=1.5$ and $k=2.5$, respectively. Upper
horizontal panel corresponds to $|T|=1$, lower to $|T|=2.$.}
 \label{ssb5T1}
  \end{figure}

At this saturation strength, only a single-solution regime persists for all
$k$ and $|T|$ values. We will pick two representative $|T|$ cases to present
the results below, $|T|=1$ and $|T|=2$, to see how the scenario differs
for relatively small versus larger intensities.
Stationary solution plots are shown in Fig.\ref{ssb5T1}.
Note that for $|T|=2$, only the case $k=2.5$ corresponds to a considerable
transmission coefficient (cf.\ Figs.\ \ref{u00001}, \ref{vv1111})

\begin{figure}[!htbp]
  \begin{minipage}[h]{0.3\linewidth}
    \centering
    \includegraphics[width=\linewidth]{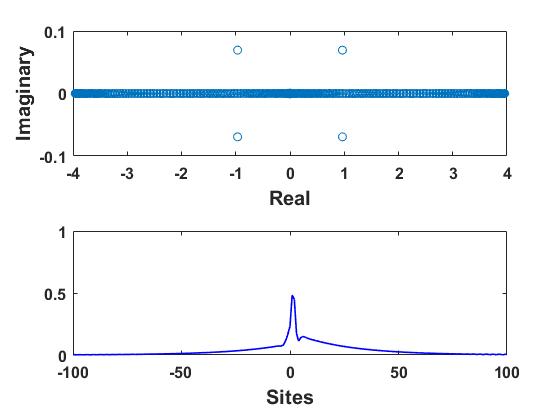}
  \end{minipage}
  \begin{minipage}[h]{0.3\linewidth}
    \centering
    \includegraphics[width=\linewidth]{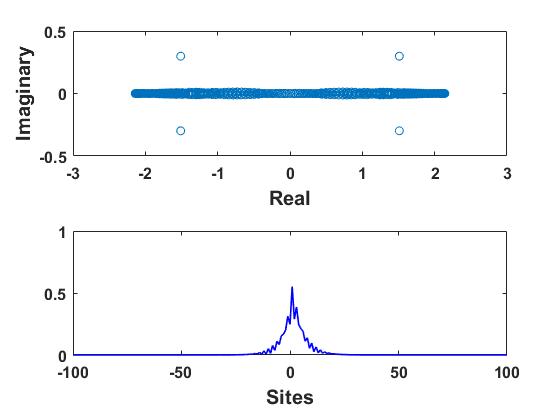}
  \end{minipage}
  \begin{minipage}[h]{0.3\linewidth}
    \centering
    \includegraphics[width=\linewidth]{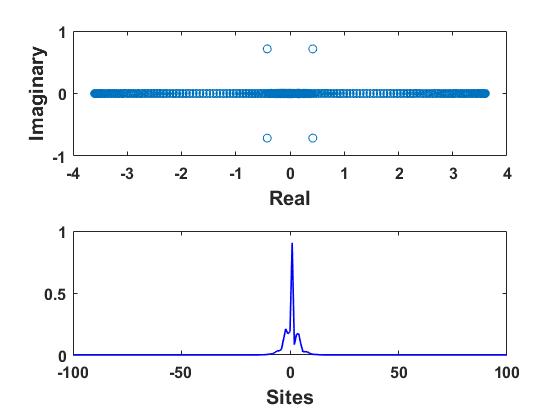}
  \end{minipage}
  \\
  \begin{minipage}[h]{0.3\linewidth}
    \centering
    \includegraphics[width=\linewidth]{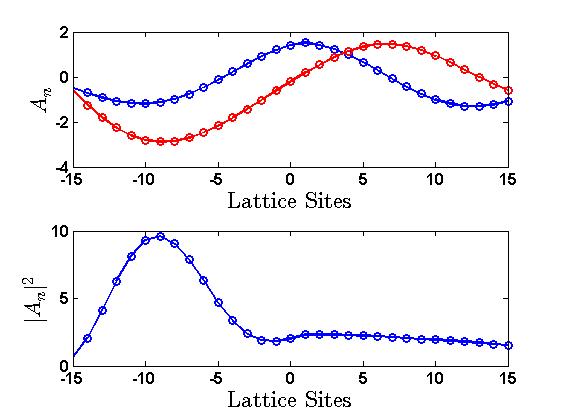}
  \end{minipage}
  \begin{minipage}[h]{0.3\linewidth}
    \centering
    \includegraphics[width=\linewidth]{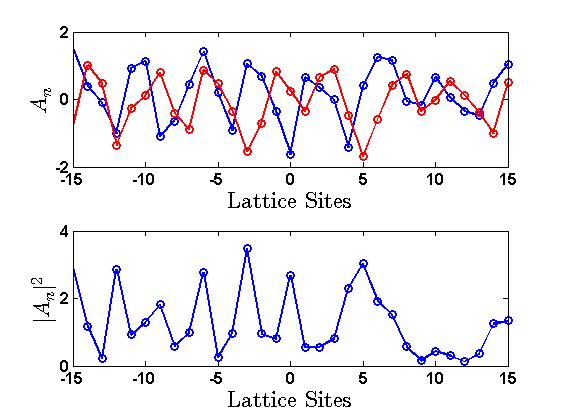}
  \end{minipage}
  \begin{minipage}[h]{0.3\linewidth}
    \centering
    \includegraphics[width=\linewidth]{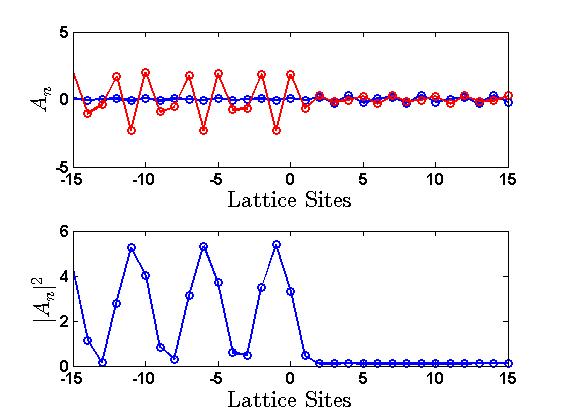}
  \end{minipage}
  \caption{Top two horizontal panels: Stability analysis for $|T|=1$ and
$\beta=0.5$, with eigenvalues at top and most unstable eigenvectors at
second row. Left, middle and right plots correspond to $k=0.2$, $k=1.5$ and
$k=2.5$, respectively. Lower two horizontal panels: snap shots of time
propagation for the corresponing values of $k$ (at time 200 for $k=0.2$ and
100 for $k=1.5, 2.5$). Third row: real (blue) and imaginary (red) parts;
fourth row:  $|A_n|^2$.}
 \label{b5evalsK02SSR}
  \end{figure}

The stability analysis with corresponding examples of unstable time
propagation for the case of $|T|=1$ and the three sample $k$ values is
presented in Fig.\ref{b5evalsK02SSR}.
Comparing with Figs.\ \ref{b01evals}-\ref{tpropb01sol1}, we note that the
scenario is very similar to that of the first solution for the same value of
$|T|$ when $\beta=0.01$; the main qualitative difference is seen for
$k=2.5$ where the instability now is oscillatory and slightly weaker. Thus,
we may conclude that as long as the amplitudes at the dimer sites are
moderate, which will typically be the case when $|T|$ is relatively small and
transmission coefficient $t$ is relatively large, the scenario in the regime
of medium saturation is qualitatively similar to that for the first
solution branch for weaker saturability.

\begin{figure}[!htbp]
  \begin{minipage}[h]{0.3\linewidth}
    \centering
    \includegraphics[width=\linewidth]{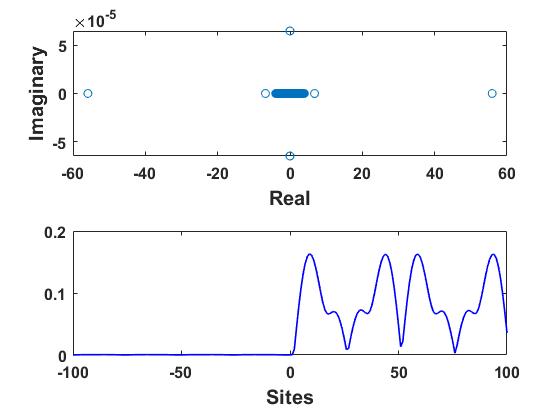}
  \end{minipage}
  \begin{minipage}[h]{0.3\linewidth}
    \centering
    \includegraphics[width=\linewidth]{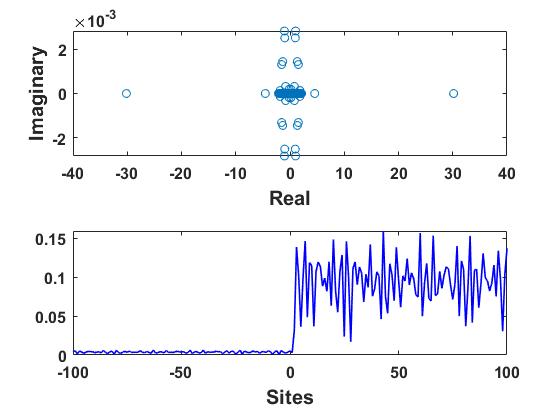}
  \end{minipage}
  \begin{minipage}[h]{0.3\linewidth}
    \centering
    \includegraphics[width=\linewidth]{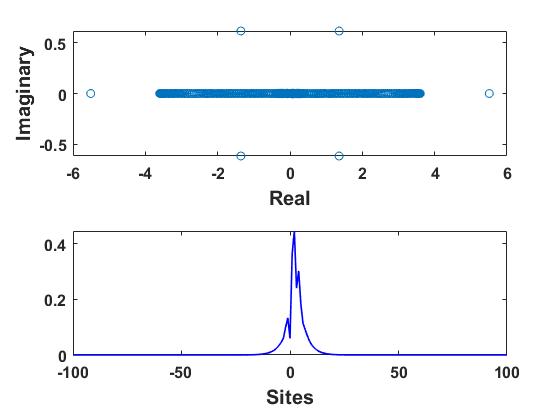}
  \end{minipage}
  \\
  \begin{minipage}[h]{0.3\linewidth}
    \centering
    \includegraphics[width=\linewidth]{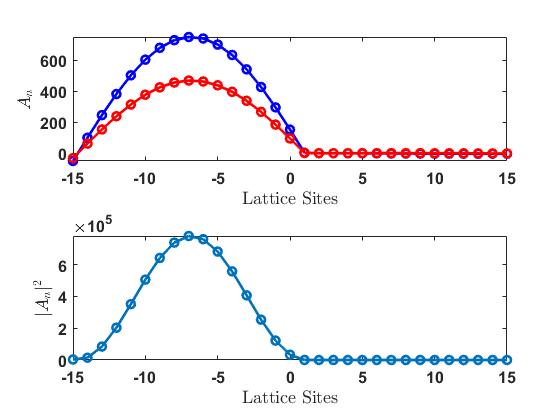}
  \end{minipage}
  \begin{minipage}[h]{0.3\linewidth}
    \centering
    \includegraphics[width=\linewidth]{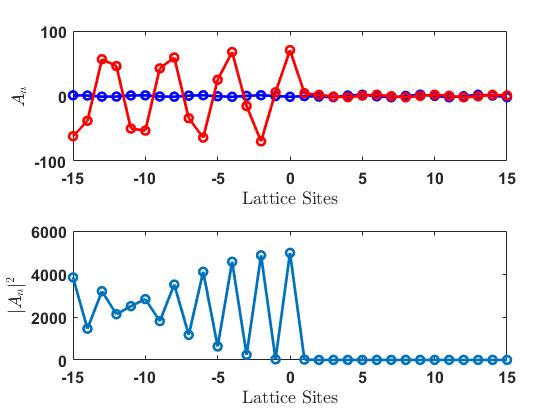}
  \end{minipage}
  \begin{minipage}[h]{0.3\linewidth}
    \centering
    \includegraphics[width=\linewidth]{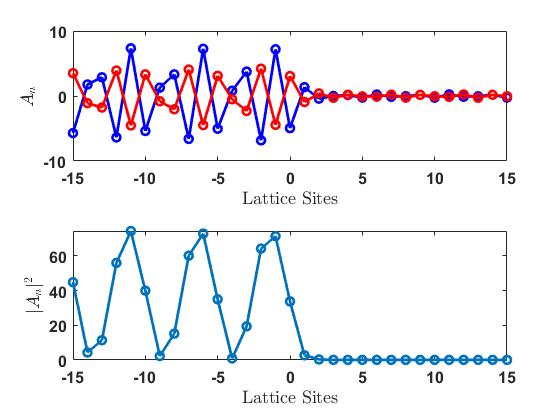}
  \end{minipage}
   \caption{Similar as Fig.\ \ref{b5evalsK02SSR} but for $|T|=2$. The
snapshots for time propagation plots are all at time 300.
}
 \label{b5evalsK15SSR}
  \end{figure}

  The case of various sample $k$ at higher intensities ($|T|=2$)
is shown in Fig.\ \ref{b5evalsK15SSR}.
For $k=0.2$ (left Fig.\ \ref{b5evalsK15SSR}) and $k=1.5$ (middle
Fig.\ \ref{b5evalsK15SSR}) the solution is now stable (but with very small
transmission coefficient), while for $k=2.5$ (right Fig.\ \ref{b5evalsK15SSR})
the instability scenario is
similar as above for $|T|=1$, resulting after some time in an almost
stationary transmission with considerably smaller amplitudes at the
transmitted side of the dimer ($|A_2|\approx 0.57$ at time 300).

Summarizing this section, we note some similarities and differences
compared to previously reported stability
results for pure on-site nonlinearities
\cite{lepri,jd}. As in \cite{lepri,jd}, we observe that the
exactly stationary propagating solutions are unstable in regimes where the
transmission coefficient is significant. Stationary waves having small
transmission
coefficient appear typically as stable as noted in \cite{lepri,jd}, except
in the multisolution regimes where the third
solution branch (which has no counterpart neither for systems with pure on-site
nonlinearity, nor with unsaturated inter-site nonlinearities) is generally
unstable, and the second solution branch destabilizes close to its
bifurcation with the third solution. As long as the instability is weak,
it is typically
oscillatory (complex eigenvalues) as originally reported in \cite{lepri}.
In regimes of stronger instabilities eigenvalues become purely imaginary, as
also seen for the saturable on-site nonlinarity in \cite{jd}. However, a
major difference is that in all cases reported in \cite{lepri,jd} for
on-site nonlinearities, the
instability resulted in a trapped, localized defect mode at the central
nonlinear sites. As seen above, none of the here considered instability regimes
resulted in any significant trapping at the dimer. Thus, it appears that
inter-site nonlinearities generally
counteract the creation of a localized dimer mode. Another difference is that
in \cite{jd}, it was concluded that instabilities generically
(for on-site saturable oligomers) appeared to transport power to the right
part of the lattice for $k>0$ (thus decreasing the power of the part
immediately to the left of the dimer). Here, we observe this scenario in some
cases (mainly for first solution branch with small $k$ when instability is
oscillatory, and third branch  at larger $k$), while in other cases the
scenario is opposite with a decrease of power at the right side
(e.g., for large $k$ in the single-solution regime and for first solution
in multi-solution regime, and for small $k$ for third solution). Thus, even
though the transmission coefficient for stationary transmission
in the regime of medium saturation
(Figs.\ \ref{u00001}, \ref{vv1111}) looks qualitatively similar to the
on-site nonlinearity cases, the instability-induced dynamics may be quite
different.

  \section{Propagation of an initial Gaussian}
\label{sec:Gaussian}

\begin{figure}[!htbp]
   \begin{minipage}[h]{0.4\linewidth}
    \centering
    \includegraphics[width=\linewidth]{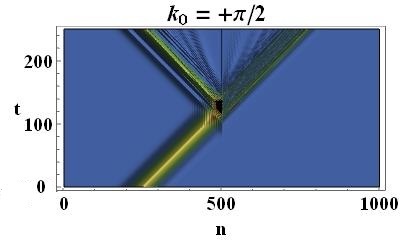}
  \end{minipage}
  \hspace{0.5cm}
  \begin{minipage}[h]{0.4\linewidth}
    \centering
   \includegraphics[width=\linewidth]{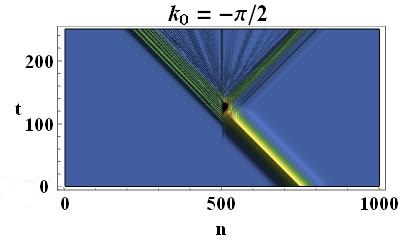}
  \end{minipage}
 \begin{minipage}[h]{0.4\linewidth}
    \centering
    \includegraphics[width=\linewidth]{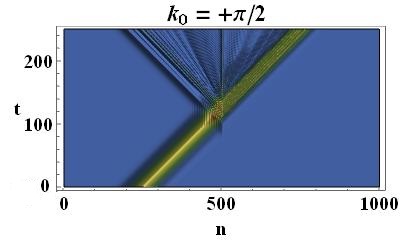}
  \end{minipage}
  \hspace{0.5cm}
  \begin{minipage}[h]{0.4\linewidth}
    \centering
    \includegraphics[width=\linewidth]{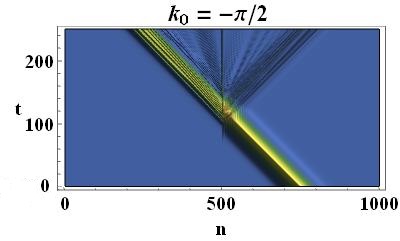}
  \end{minipage}
   \caption{Color plots of $|A_n(t)|^2$ as a function of $n$ (lattice sites)
and $t$ (time), with the Gaussian initial data of Eq.\ \eqref{gwp}, saturation
$\beta=0.05$ (upper horizontal panel) and
$\beta=0.5$ (lower horizontal panel); wavenumber $k_0=\pm\frac{\pi}{2}$,
and all other parameters as before. Left (right) vertical panel corresponds to
the cases with left (right) incidence. The dimer is located at sites
501 and 502 in a 1000 site lattice.
}
 \label{gwpb00001}
   \end{figure}

  As an example different from the stationary plane wave solutions we
investigate, by direct numerical integration of
Eq.\ \eqref{dynamical}, the time propagation of an initial Gaussian wavepacket
through the chain with the dimer defect having saturated inter-site
nonlinear interactions between the two dimer sites.
The initial Gaussian data is
\bea{}
A_n(0)= I~ \textmd{exp}\left[-\frac{\left(n-n_0\right)^2}{w^2}+ik_in\right] ,
\label{gwp}
\eea
where $I$ and $p$ are the amplitude and width of the initial wave-packet taken
to be $\sqrt{3}$ and $56$, respectively. Typical results for this initial
condition with $|k_0|=\pi/2$ (corresponding
to maximum propagation speed and minimum dispersion) in the regimes of low
($\beta=0.05$) and medium ($\beta=0.5$) saturation are shown in
Fig.\ \ref{gwpb00001}. (Results in the ultra-low saturation regime are
very similar to those for $\beta=0.05$.)

The transmission coefficient for the wavepacket is defined
as in \cite{11} to be the ratio of transmitted power to the initial input
power. For the case of right propagating signal, it is given by
  \bea{}
  \tau_+=\frac{\sum_{i>\frac{n}{2}+2}|A_i(t_f)|^2}{\sum_{i<\frac{n}{2}+1}|A_i(0)|^2} ,
  \label{tcR}
  \eea
where $n$ is the total number of sites, the dimer is located at
sites $n/2+1$ and $n/2+2$, and $t_f$ the final time for the
numerical integration. Equation \eqref{tcR} can be understood by a direct
analogy with Eq.\ \eqref{t}. For $\beta=0.05$,
the transmission coefficient for the left
incidence (right propagating case; top left plot in Fig.\ \ref{gwpb00001})
is found to be $\tau_+=0.474$. In an analogous way, the transmission
coefficient for the right incidence (left propagating) case depicted in
the top right plot in Fig.\ \ref{gwpb00001} turns out to be
$\tau_-=0.603$. The rectifying factor is computed by the formula
$f=\frac{\tau_+-\tau_-}{\tau_++\tau_-}$, in analogy with Eq.\ \eqref{rf},
which in this case is found to be $f=-0.120$. The distinct left and right
transmission coefficients reflect the fact that the parity symmetry of the
dimer defect is indeed broken.
The lower horizontal panel in Fig.\ \ref{gwpb00001} corresponds to the regime
of medium $\beta$ strength ($\beta=0.5$). The transmission coefficients are
found to be $\tau_+=0.797$ and $\tau_-=0.819$, and the corresponding
rectifying factor is found to be $f=-0.013$.
Thus, increasing saturation implies that the transmission increases but the
asymmetry decreases significantly, as was also found in previous
works for on-site saturability \cite{Erik,wasay3}. This is also consistent
with results for the stationary transmission, e.g., by comparing middle left
plot in Fig.\ \ref{k151}  and middle plot in Fig.\ \ref{vv1111}.
For $\beta=0.05$ and $|k|$ close to $\pi/2$ (Fig.\ \ref{k151})
there is essentially no transmission
with $|T|^2 > 2$, and thus the peak intensity of the Gaussian with $I^2 = 3$
cannot be transmitted, while for $\beta=0.5$ (Fig.\ \ref{vv1111}) there
is almost complete transmission around $|T|^2 = 2.5$, allowing for a larger
portion of the Gaussian to be transmitted. On the other hand, the main
transmission around $|T|^2 = 2.5$ for $\beta=0.5$ is almost symmetric for
left- and right-propagation, while for  $\beta=0.05$ the stationary
transmission for right-propagation (blue curve) dips sharply around
$|T|^2 = 1.8$, while for left-propagation (red curve) the dip appears later,
around $|T|^2 = 2$. Thus, a considerable part of the Gaussian  with $I^2 = 3$
may be transmitted to the left but not to the right. (Weak instabilites of
the stationary transmission modes described in previous section will
not significantly affect the transmission of a rapidly moving and
not too wide Gaussian, if the time for the Gaussian to pass the dimer will
be shorter than the time for the instability to develop.)

It should also be noted from Fig.\ \ref{gwpb00001} that, in addition to
partial transmission/reflection of the Gaussian and creation of
small-amplitude radiation waves, a rather small part will remain trapped
at the dimer sites. For the cases shown in Fig.\ \ref{gwpb00001}, we find
at time $t_f=250$ the trapped intensity $P_{trap}\equiv |A_{501}|^2+|A_{502}|^2$ to
be:  $P_{trap}(\beta=0.05, k=+\pi/2)=0.77$,
$P_{trap}(\beta=0.05, k=-\pi/2)=2.17$,
$P_{trap}(\beta=0.5, k=+\pi/2)=2.79$
$P_{trap}(\beta=0.5, k=-\pi/2)=2.29$.
Thus, in these cases considerably less trapping appears in the case when the
main part is reflected, compared to when it is mainly transmitted. This is
opposite to the example considered in \cite{11}, where trapping was enhanced
when the main part was reflected. We may also note that for on-site
saturabilites, increasing saturation strength typically decreases trapping
\cite{Erik}, while we here observed an opposite tendency (which might be
intuitively understood as increasing the saturability of the inter-site
nonlinearity implies that the on-site nonlinearity will be relatively more
important, and trapping is in general mainly associated with on-site
nonlinearities).


Note also that the transmission coefficient typically is largest
when the incoming Gaussian first encounters the dimer site with smallest
$|V_2|$, i.e., the site where the deviation from the linear chain with
zero on-site potential is smallest. Intuitively this seems reasonable, and an
analogous remark was made for a saturable on-site potential in \cite{jd}.
We checked a few other parameter values and obtained analogous results.
Keeping $V_1=-2.625$ and changing the on-site amplitude on the right
dimer site to $V_2=-1.875$ (i.e., decreasing its magnitude), we obtained
for $\beta=0.05$ that $\tau_+=0.594$ and $\tau_-=0.746$ with
rectifying factor $f=-0.113$, i.e., transmission increases while rectification
remains almost the same (and with the same sign). On the other hand, changing
the on-site amplitude at the right dimer site to $V_2=-3.125$
(i.e., increasing its magnitude to have $|V_2|>|V_1|$) resulted in
 $\tau_+=0.722$ and $\tau_-=0.542$ with $f=+0.143$. This means that the
diode-like transmission in the low-saturation regime, as expected, gets
reversed when $|V_2|>|V_1|$.

\section{Conclusions}
\label{sec:conclusions}

The main aim of this work has been to provide a clear and comprehensive
description of qualitatively novel effects that appear in the
transmission scenario of a DNLS-type dimer when nonlinear coupling between
the dimer sites is taken into account, in addition to a standard onsite
(cubic) nonlinearity. For the generality of the description, we considered
a saturable intersite nonlinearity, with a parameter $\beta$ interpolating
between a purely cubic intersite + onsite nonlinearity at $\beta=0$, and
a cubic pure onsite nonlinearity at large $\beta$. A major novel result is
that, in contrast to the commonly studied cases with pure onsite nonlinearity,
the transmission coefficient for stationary transmission is in general no
longer a single-valued function
of the transmitted intensity $|T|$, as the standard backward transfer map in
regimes of small and moderate
$|T|$ and low saturation will have three distinct solutions
(two in the case of strictly zero saturation). These solutions
differ in their relative phase shift across the dimer.
As the saturation
increases, solutions disappear through bifurcations at critical saturation
strengths (depending on the wave number), leaving a single solution
branch in regimes of medium and strong saturation.

We analyzed numerically the transmission coefficient of the three different
branches, and showed how these merged into the single-valued (as a function
of $|T|$) transmission picture, known from previous works, as saturation
strength increased. We also performed a linear stability analysis
of the stationary
scattering solutions. In low-saturation regimes with three solution branches,
one branch, with very small transmission, exhibited strong instabilities,
while from the other two branches solutions with small transmission
coefficients were typically stable, and those with larger transmission
exhibited rather weak instabilites, growing larger when approaching
transmission peaks. Qualitatively the results for the latter two
branches agree with previous work for pure onsite DNLS. Studying by direct
numerical simulations the effect of the instabilities on the transmission, two
notable differences to previous works were found as a result of the
intersite nonlinearities: (i) we did not observe any
significant trapping at the dimer sites; (ii) while previous works always
found transport of power towards the transmission side, in some regimes we
instead found transport of power towards incoming/reflected side.

We also analyzed the left/right asymmetries in the transmission coefficient
with different linear onsite potentials at the dimer sites, both for
stationary plane waves and for
rather wide and rapidly moving Gaussian excitations with amplitiude near the
main transmission threshold. In addition to shifting the location of
transmission peaks, the multiple transmission branches for low saturability
leads to a novel rectification effect for stationary transmission, where a
transmission peak in one direction, for a given branch and a given $|T|$,
may correspond to non-existence of solutions in the opposite propagation
direction for this branch. For the Gaussian propagation, we found that
increasing saturation strength typically would increase the transmission
coefficient but decrease the rectifying factor, as the stationary transmission
spectrum also became broader and more symmetric for wave vectors close to
$\pi/2$.

Finally, we also comment on possible applications of our work.
Our choice of model arose from the interest in studying the gradual transition
from a system with non-saturated
to saturated intersite nonlinearities, keeping onsite nonlinearities
non-saturated. The saturability is typically in the form
appearing from photovoltaic-photorefractive materials, which may be a suitable
class of systems where the phenomenology of multichannel asymmetric
transmission described here may be observed. A relevant topic for future
research would be to perform a similar analysis for the type of saturable
nonlinear couplings proposed in \cite{Hadad17}, with direct application to
electric circuit ladders \cite{Hadad18}.

\section*{Acknowledgments}

M.A.W would like to thank Jennie D'Ambroise for several fruitful
discussions and to Byoung S. Ham for the facilities. M.A.W acknowledges financial support by the ICT R\&D program of MSIT/IITP (1711073835: Reliable crypto-system standards and core technology development for secure quantum key distribution network) and GRI grant funded by GIST in 2018. M.J. thanks Erik Johansson for discussions during his thesis work
\cite{Erik}, which served as a source of inspiration for the present work.


 \bibliographystyle{unsrt}

\appendix*



\section{Slightly increasing/decreasing $V_2$ at $\beta=0.01$}
\label{AppB}
\begin{figure}[!htbp]
  \begin{minipage}[h]{0.34\linewidth}
    \centering
    \includegraphics[width=\linewidth]{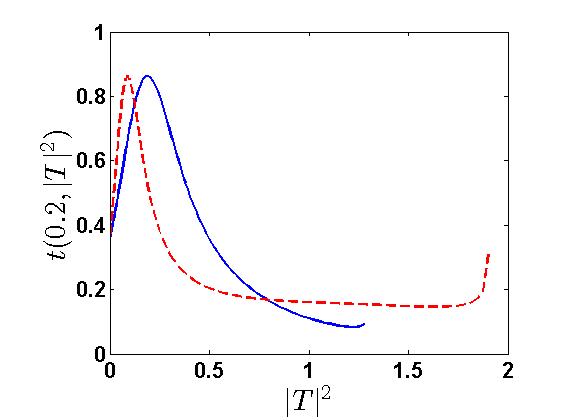}
  \end{minipage}
  \hspace{-0.7cm}
   \begin{minipage}[h]{0.34\linewidth}
    \centering
    \includegraphics[width=\linewidth]{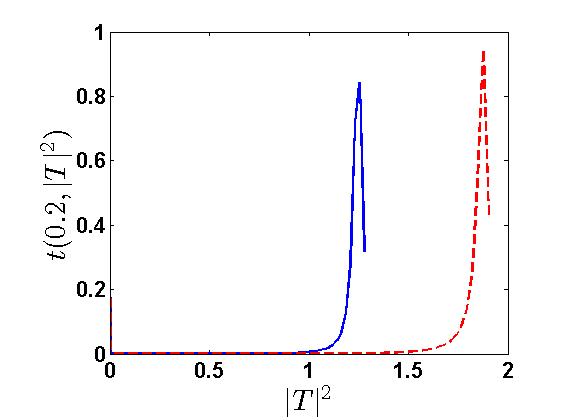}
  \end{minipage}
  \hspace{-0.7cm}
   \begin{minipage}[h]{0.34\linewidth}
    \centering
    \includegraphics[width=\linewidth]{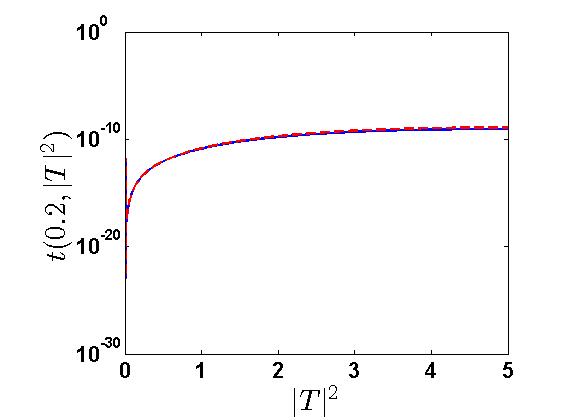}
  \end{minipage}
  \\
  \begin{minipage}[h]{0.34\linewidth}
    \centering
    \includegraphics[width=\linewidth]{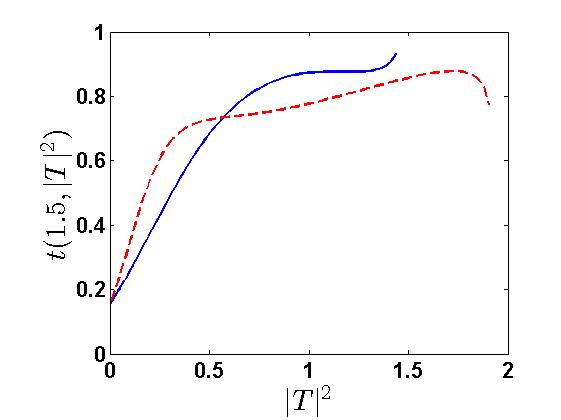}
  \end{minipage}
  \hspace{-0.7cm}
   \begin{minipage}[h]{0.34\linewidth}
    \centering
    \includegraphics[width=\linewidth]{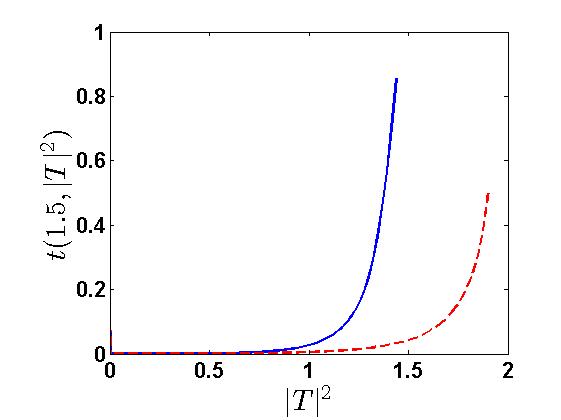}
  \end{minipage}
  \hspace{-0.7cm}
   \begin{minipage}[h]{0.34\linewidth}
    \centering
    \includegraphics[width=\linewidth]{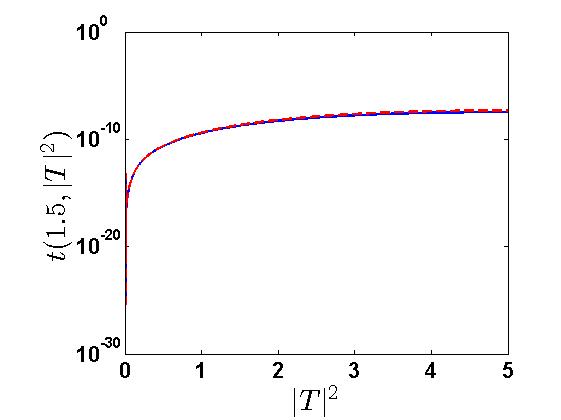}
  \end{minipage}
  \\
  \begin{minipage}[h]{0.34\linewidth}
    \centering
    \includegraphics[width=\linewidth]{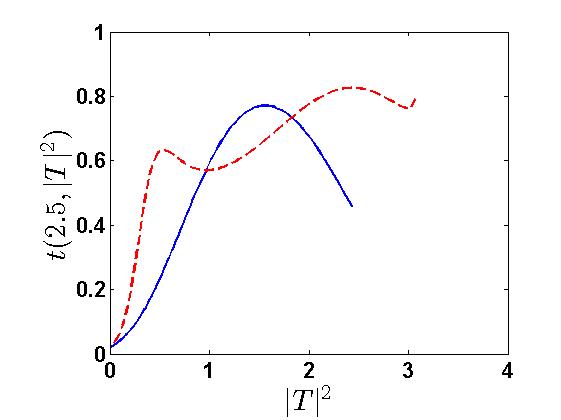}
  \end{minipage}
  \hspace{-0.7cm}
   \begin{minipage}[h]{0.34\linewidth}
    \centering
    \includegraphics[width=\linewidth]{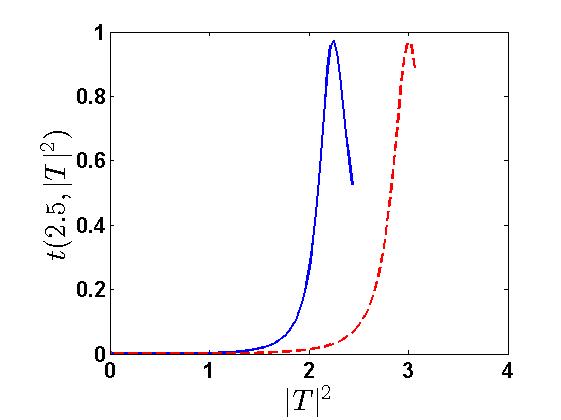}
  \end{minipage}
  \hspace{-0.7cm}
   \begin{minipage}[h]{0.34\linewidth}
    \centering
    \includegraphics[width=\linewidth]{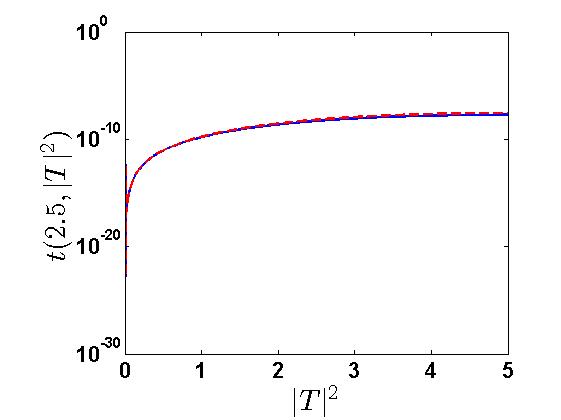}
  \end{minipage}

   \caption{Same as Fig.\ref{vv1}, except $V_2$ has been increased to -1.8750.}
 \label{V2eL}
  \end{figure}

To exemplify how the stationary transmission scenario for left- and
right-propagating waves changes when the linear
onsite potential on the dimer sites is varied, we consider the
transmission curves at $\beta=0.01$ and first increase $V_2$ by
setting $V_2=V^{(0)}(1-5\varepsilon)$($=-1.8750$), where $\varepsilon$ is the
asymmetry held at $\varepsilon=0.05$, and $V^{(0)}=-2.5$. Note that $V_1$
remains at $V_2=V^{(0)}(1+\varepsilon)(=-2.6250)$, as in the whole paper.
The transmission curves corresponding to all three solution branches for all
three
representative $k$ are shown in Fig.\ \ref{V2eL}.
Compared with the ones in Fig.\ \ref{vv1}, where $V_2$ was kept at
$V_2=V^{(0)}(1-\varepsilon)(=-2.3750)$,  we first note that increasing $V_2$
shrinks the multi-solution regime for the right-propagating waves
(blue curves) but does not change the existence regimes for left-propagating
waves (red curves). This is simply a consequence of the fact that \eqref{argT}
contains $V_2$ but not $V_1$.
%
The transmission coefficent however depends on both $V_1$ and $V_2$ for
both directions of propagation, and is overall decreased as
$|V_1-V_2|$ increases. Thus, rectification effects increase, but to the
price of narrower transmission peaks and lower total transmission.

\begin{figure}[!htbp]
  \begin{minipage}[h]{0.34\linewidth}
    \centering
    \includegraphics[width=\linewidth]{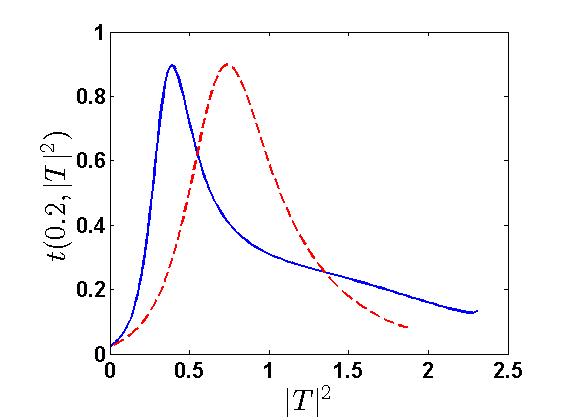}
  \end{minipage}
  \hspace{-0.7cm}
   \begin{minipage}[h]{0.34\linewidth}
    \centering
    \includegraphics[width=\linewidth]{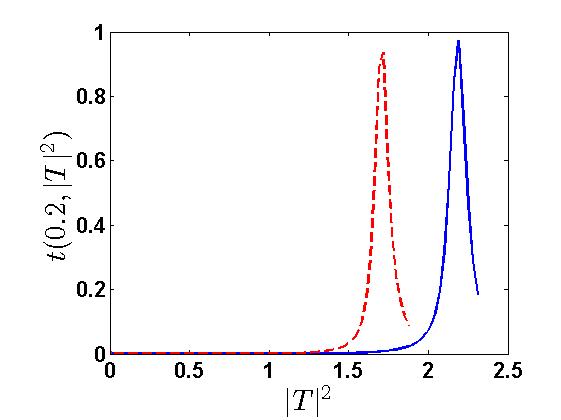}
  \end{minipage}
  \hspace{-0.7cm}
   \begin{minipage}[h]{0.34\linewidth}
    \centering
    \includegraphics[width=\linewidth]{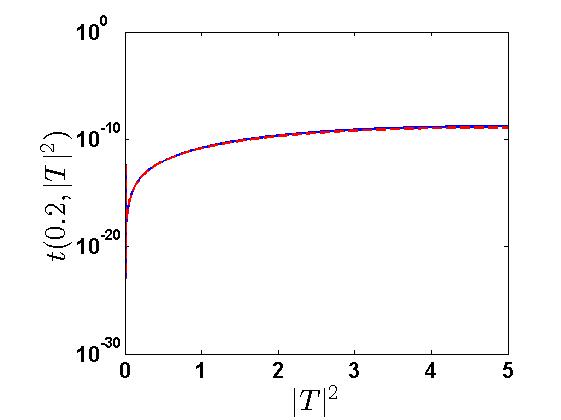}
  \end{minipage}
  \\
  \begin{minipage}[h]{0.34\linewidth}
    \centering
    \includegraphics[width=\linewidth]{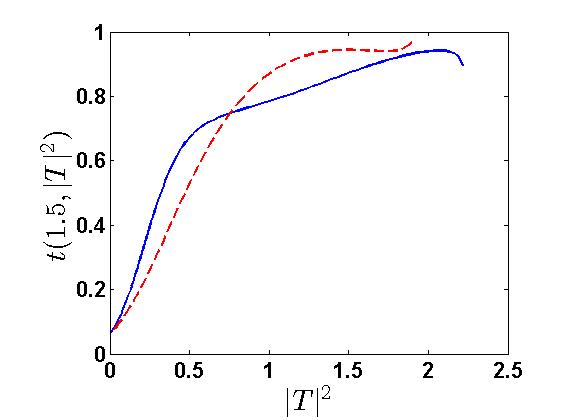}
  \end{minipage}
  \hspace{-0.7cm}
   \begin{minipage}[h]{0.34\linewidth}
    \centering
    \includegraphics[width=\linewidth]{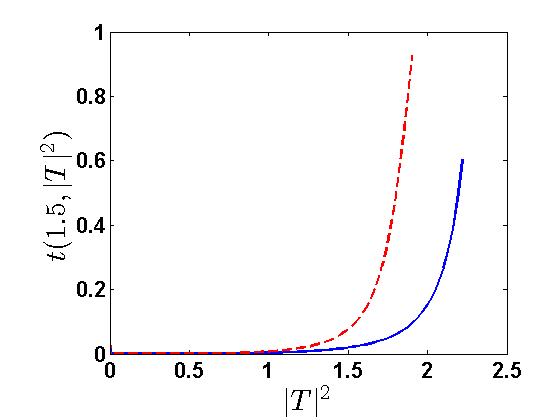}
  \end{minipage}
  \hspace{-0.7cm}
   \begin{minipage}[h]{0.34\linewidth}
    \centering
    \includegraphics[width=\linewidth]{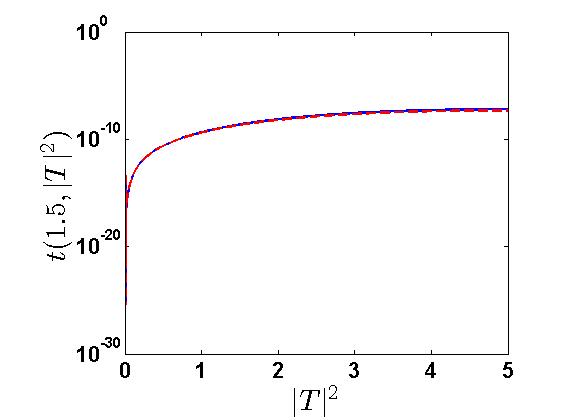}
  \end{minipage}
  \\
  \begin{minipage}[h]{0.34\linewidth}
    \centering
    \includegraphics[width=\linewidth]{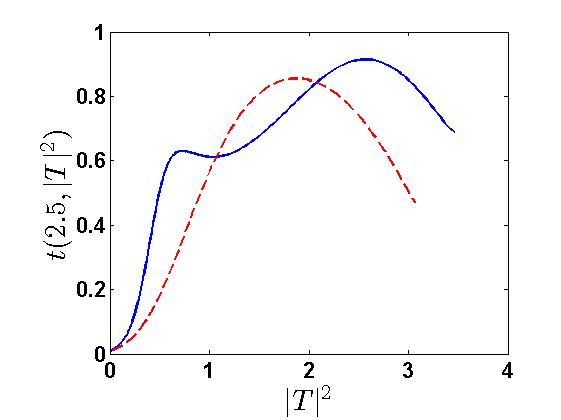}
  \end{minipage}
  \hspace{-0.7cm}
   \begin{minipage}[h]{0.35\linewidth}
    \centering
    \includegraphics[width=\linewidth]{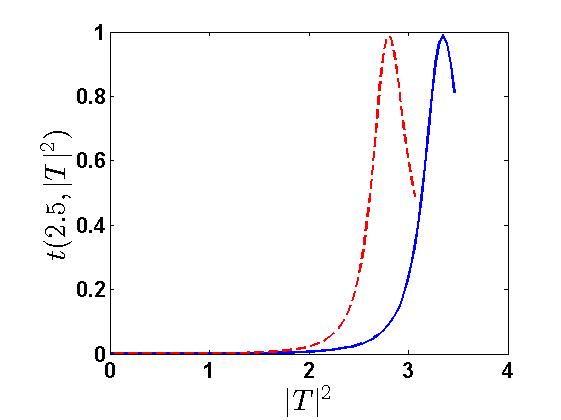}
  \end{minipage}
  \hspace{-0.7cm}
   \begin{minipage}[h]{0.34\linewidth}
    \centering
    \includegraphics[width=\linewidth]{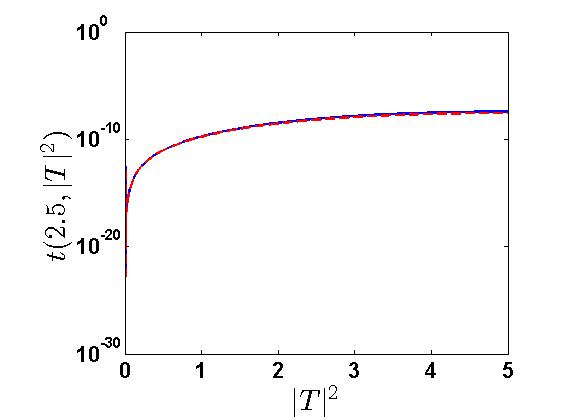}
  \end{minipage}

   \caption{Same as Fig.\ \ref{vv1}, except $V_2$ has been decreased
to -3.1250.}
 \label{V2eS}
  \end{figure}

 The case of decreasing $V_2$ to
 $V_2=V^{(0)}(1+5\varepsilon)(=-3.1250)$ is shown in Fig.\ \ref{V2eS}.
We note that decreasing $V_2$ results in the persistence of multi-solutions
for a longer stretch of intensities for right-propagating waves. In addition to
this, since now $|V_2| > |V_1|$ the main transmission scenario gets
reversed between the left and right incidence, as compared to Fig.\ \ref{vv1}.
Generally, the transmission regime appears wider when the incoming wave
first hits the site with smallest magnitude of the linear on-site potential.

\end{document}